\documentclass[namedreferences]{solarphysics}

\usepackage[hyperref,optionalrh]{spr-sola-addons}
\usepackage{graphicx}        
\usepackage{color}           
\usepackage{breakurl}        

\chardef\us=`\_

\begin{document}

\begin{article}
\begin{opening}

\title{Shaken and stirred: When Bond meets Suess-de Vries and Gnevyshev-Ohl}

\author[addressref={aff1},corref,email={F.Stefani@hzdr.de}]{\inits{F.}\fnm{F.}~\lnm{Stefani}}\sep
\author[addressref={aff2,aff3}]{\inits{A.}\fnm{R.}~\lnm{Stepanov}}\sep
\author[addressref={aff1}]{\inits{T.}\fnm{T.}~\lnm{Weier}}\sep
\address[id=aff1]{Helmholtz-Zentrum Dresden -- Rossendorf, Bautzner Landstr. 400,
D-01328 Dresden, Germany}
\address[id=aff2]{Institute of Continuous Media 
Mechanics,1 Acad. Korolyov str., 614013 Perm, Russia}
\address[id=aff3]{Perm National Research Polytechnic 
University, Komsomolskii av. 29, 614990 Perm, Russia }

\runningauthor{F. Stefani {\it et al.}}
\runningtitle{Shaken and stirred}

\begin{abstract} We argue that the most prominent 
temporal features of the solar dynamo, in particular the 
Hale cycle, 
the Suess-de Vries cycle (associated with 
variations of the Gnevyshev-Ohl rule), Gleissberg-type cycles, 
and grand minima can be self-consistently explained 
by double synchronization with the 11.07-years periodic 
tidal forcing of the Venus--Earth--Jupiter system and the
(mainly) 19.86-years periodic motion of the Sun around
the barycenter of the solar system.
In our numerical simulation, grand minima, and clusters thereof,  
emerge as intermittent and non-periodic events on millennial 
time scales, very similar to the series of Bond events which were 
observed throughout the Holocene and the last glacial period. 
If confirmed, such an intermittent transition to chaos 
would prevent any long-term prediction of solar activity, 
notwithstanding the fact that the shorter-term Hale and 
Suess-de Vries cycles are clocked by planetary 
motion. 
\end{abstract}
\end{opening}
\section{Introduction}

Thanks to the seminal work of Gerard Bond and his collaborators, we 
now have overwhelming evidence that a significant component of 
sub-Milankovich 
climate variability occurs in certain 1-3-kyears ``cycles''
of abrupt changes of the North Atlantic's surface 
hydrography \citep{Bond1997,Bond1999}, and that 
those {\it Bond events} are closely related to corresponding 
variations in solar output, evidenced by measurements of 
the cosmogenic radionuclides 
$^{14}$C and $^{10}$Be \citep{Bond2001}. 
While originally identified for the Holocene \citep{Bond1997} 
from certain ice drift proxies (in particular volcanic 
glass from Iceland and 
hematite-stained grains from East Greenland,
found in deep-see sediments), 
many similar events were later revealed by 
\cite{Bond1999} in the last glacial period, too. 
Viewed from this angle, the little ice
age, comprising in particular 
the Sp\"orer and  Maunder grand minima, 
appears just as the latest link in the 
chain of Bond events, and the temperature increase 
since the end of the Dalton minimum as 
a rebound from those frosty times.  

\begin{figure}[t]
  \centering
  \includegraphics[width=0.8\textwidth]{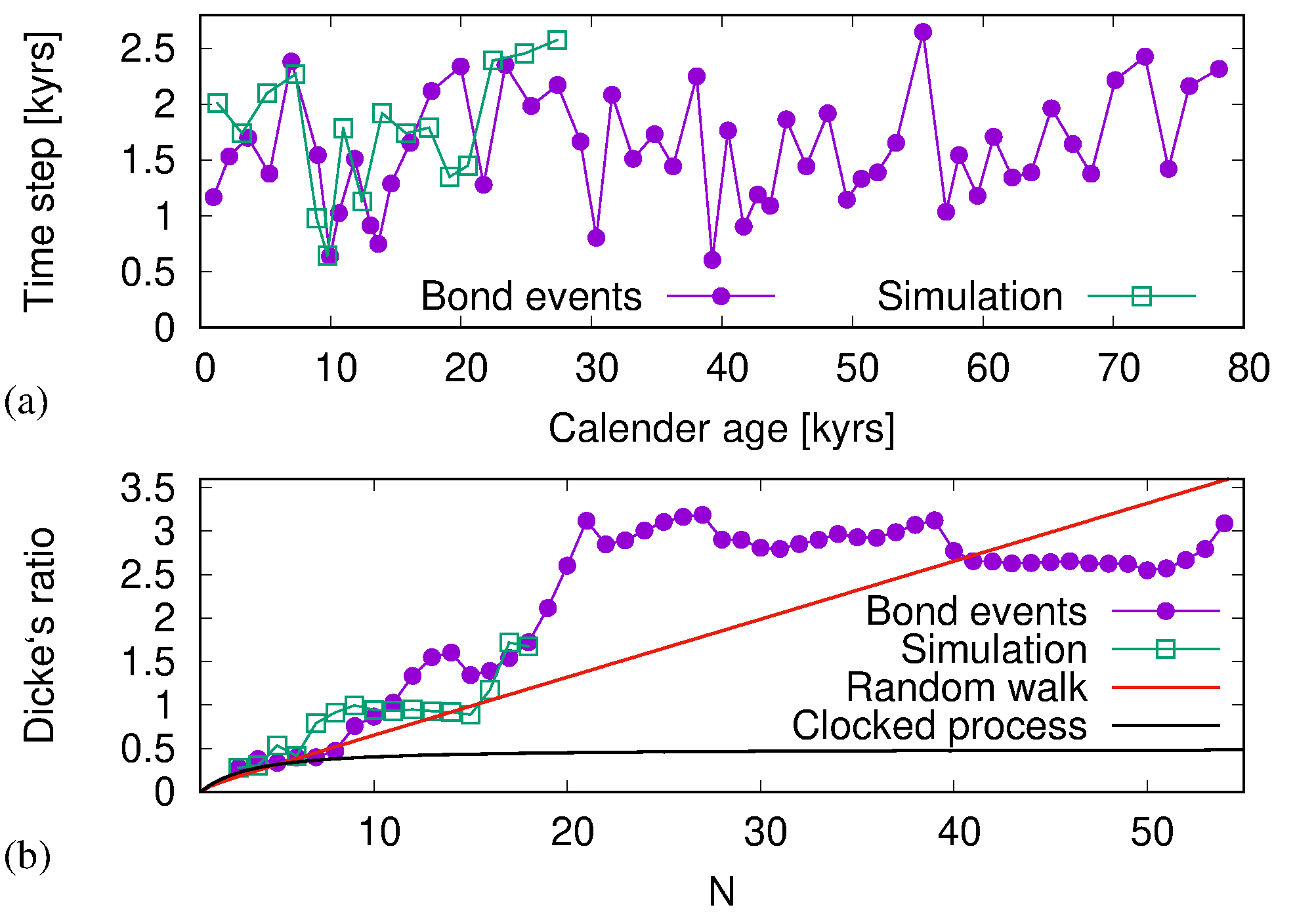}
  \caption{(a) Violet full dots are time separations of the Bond events for the last 
  80 kyears, as inferred from Figure 6(c) of \cite{Bond1999} for 
  the data of the deep-sea sediment core VM23-81. 
  Green empty squares denote the corresponding time separations 
  of numerically simulated ``Bond events'' for the parameters
   $\omega_0=10000$, $\alpha^c_0=15$, $\alpha^p_0=50$
  $q^p_{\alpha}=0.2$, $q^c_{\alpha}=0.8$, $D=0.05$ and 
  $\kappa(t)=0.6+0.2 m(t)$). (b) Dependence of Dicke's ratio on $N$
  for the two time series from (a), together with the theoretical 
  curves for a random walk
  and a clocked process. Note that the horizontal positions of the
  individual points slightly differ between (a) and (b) since 
  the abscissa in (a) shows the centre point for the time separation, 
  while the abscissa in (b) indicates the number of Bond events taken 
  into account.
  }
  \label{fig:bond_and_sim}
\end{figure}

In solar physics, similar variations with 
time scales of 1-3 kyears are 
usually discussed under the notion 
{\it Eddy cycle}
and  {\it Hallstadt cycle} 
\citep{Steinhilber2012,Abreu2012,Soon2014,Scafetta2016,Usoskin2016}. 
Yet, some caution seems to be appropriate
when stretching the very concept of ``cycles'' from the
decadal  (Schwabe, Hale) to the millennial time scale, 
in particular when
the underlying $^{14}$C and $^{10}$Be 
data bases have typical durations of only 10\,kyears, 
or just slightly longer \citep{Kudryavtsev2020}.
To gain more insight into the statistics of those
``cycles'', we make here the plausible assumption that the 
established link 
between solar activity and Bond events, 
with correlation coefficients of around 0.5 during the 
Holocene \citep{Bond2001}, 
extends also to the  last glacial.
With this proviso, we re-plot in Figure \ref{fig:bond_and_sim}(a)  
the data of time separations (or waiting times)
between the 54 Bond events as identified over the last 80 kyears, 
which we have drawn from  
Figure 6(c) of \cite{Bond1999}. What we observe then is a  
broad range of time separations between 600 years and 2600 years, 
with a mean value of 1469 years and  a standard deviation of 514 years,
according to 
\cite{Bond1999}. However, when focusing 
only on the first eight time 
separations in the Holocene, located chiefly around 1500 years, 
2400 years and 600 years, it comes as no surprise
to find similar periods in Fourier or wavelet analyses 
\citep{Mayewski1997,Dima2009,Soon2014}. Special attention on 
the mean 1470-years cycle, and even speculations about 
its origin from a coincidence of 
17 Gleissberg and 7 Suess-de Vries cycles \citep{Braun2004},  
seem to be justified in this case.

If, on the contrary, we consider the entire series of Bond 
events over 80 kyears, we get the same impression as 
 \cite{Usoskin2007} who had argued 
that the ``occurrence of grand minima/maxima is driven not 
by long-term cyclic variability, but by a stochastic/chaotic process.''
More quantitatively, the random walk character of this series 
is analyzed in Figure \ref{fig:bond_and_sim}(b)
which shows Dicke's ratio
$\sum_i^N  r_i^2/\sum_i (r_i-r_{i-1})^2$ 
between the mean square of the residuals $r_i$
(i.e., the distances between the actual 
Bond events and hypothetical events according to a 
linear fit to the series) 
to the mean square of the differences $r_i-r_{i-1}$
between two consecutive residuals \citep{Dicke1978}. Obviously, Dicke's 
ratio for $N$ Bond events
taken into account  
roughly approaches the theoretical random walk dependence 
$(N+1)(N^2-1)/(3(5 N^2+6N-3))$, with its limit $N/15$,
while significantly deviating
from the corresponding dependence $(N^2-1)/(2(N^2+2N+3))$ 
for a clocked process, with its limit $1/2$, 
as it had been confirmed previously
for the Schwabe cycle \citep{Stefani2019}.

Figure \ref{fig:bond_and_sim}(a) is complemented by another curve of 
16 time separations (restricted  to an interval of 30\,kyears), as 
it will come out of our 
numerical model to be described further below. Suffice it to say here 
that the average of the waiting times, and their broad distribution, 
are quite similar to those of the 54 real Bond events, and 
that Dicke's ratio points also in direction of a random walk process,
although the statistical significance of those 
mere 17 numerical ``Bond events'' is certainly not satisfactory.  

In this paper we will try to give a relatively simple and self-consistent
answer to the questions of how such an intermittent and random walk like 
behaviour comes about, and how it can be related to the much clearer 
periodicities  of the Hale, the Suess-de Vries, and 
the Gleissberg cycle(s). For this purpose, we will start from 
a rather conventional $\alpha-\Omega$-dynamo model
in form of a simple 1D PDE system (with co-latitude as the only 
spatial variable), to allow for very long simulations of appr. 
30 kyears. 
With the source terms $\alpha$ and $\Omega$ 
appropriately chosen,
this dynamo develops typical oscillation periods of 
20-30 years. After enhancing these internal ``stirring'' terms 
of the dynamo by two external ``shaking'' terms with the specific 
periods 11.07 years and 19.86 years (both being related to 
planetary motions), we will end up 
with a more or less complete reproduction of the most 
relevant temporal features of the solar dynamo. 
But beforehand, we have to clarify 
the origin of these two external ``shaking'' terms, and 
how their use in a solar dynamo model can be justified.

We start with the 11.07-years period. Although the similarity
of the Schwabe cycle and the spring-tide period 
of the Venus--Earth--Jupiter (VEJ)
system has been known for a long time
\citep{Bollinger1952,Takahashi1968,Wood1972,CondonSchmidt1975},
the precise correspondence of these two 11.07-years 
periodicities was  recognized only recently by 
\cite{Hung2007,Scafetta2012,Wilson2013,Okhlopkov2014,Okhlopkov2016}.
According to \cite{Scafetta2012}, this period, corresponding to 
4043\,days, results
from the resonance condition
\begin{eqnarray}
P_{\rm VEJ}&=& \frac{1}{2} \left[  \frac{3}{P_{\rm V}} -\frac{5}{P_{\rm E}}+\frac{2}{P_{\rm J}} \right]^{-1}
\end{eqnarray}
for the period of VEJ alignments, with the
sidereal periods $P_{\rm V}=224.701$\,days, $P_{\rm E}=365.256$\,days,
and $P_{\rm J}=44332.589$\,days for Venus, Earth and Jupiter, 
respectively.

Quite often, the action of planetary tidal forces on the Sun
is discarded in view of the tiny acceleration 
in the order of $10^{-10}$\,ms$^{-2}$ \citep{DeJager2005,Callebaut2012},
leading to a negligible tidal height 
$h_{\rm tidal} \approx
G m R^2_{\rm tacho}/(g_{\rm tacho} d^3)$ 
in the order of  1\,mm 
($m$ is the planet's mass and $d$ its distance to the Sun). 
One should note, however, that this 
tidal height translates  (by virtue of the 
virial theorem) into a non-negligible tidal flow velocity
of $v \sim (2 g_{\rm tacho} h_{\rm tidal})^{1/2} \approx 1$\,m/s, 
when taking into account the huge gravity at the tachocline of 
$g_{\rm tacho} \approx 500$\,m/s$^{2}$ \citep{Opik1972}. But even 
then it is hard to conceptualize how tidal forces could
influence the solar dynamo without employing any sort of 
amplification mechanism. One candidate for such a mechanism 
was discussed by
\cite{Wolff2010,Scafetta2012} who speculated that planetary
forces might affect the nuclear burning rate deep in the solar core
and that this effect would be promptly felt at the tachocline 
via the resulting change in  
g-mode oscillations.  While this sounds highly 
speculative, we mention here a pertinent observation 
by \cite{Kotov2020} regarding a possible influence of 
Jupiter's synodic cycle on the  
(still controversially discussed) 160\,min oscillation 
of the Sun's radius.

An entire suite of most interesting synchronization 
models was corroborated by \cite{Wilson2013}. 
First, the 11.07-years periodicity was explained in terms of  
a VEJ tidal-torque model, 
in which Jupiter plays a distinguished role by exerting a torque 
upon the periodically forming Venus-Earth tidal bulges.
Second, a gear effect was invoked to modulate
the changes in the rotation rates (as driven by the 
tidal-torque effect)
by the 19.86-year periodic Saturn-Jupiter quadratures, 
leading ultimately to a long-term modulation with 
193-year period (which will play a key role further below). 
Further derivations of a 208-years Suess-de Vries 
cycle, a 1156-years Eddy cycle, and a 2302-years Hallstatt cycle
can also be found in this highly instructive and 
recommendable paper. 
 
In \cite{Wilson2013} and in the earlier model of 
\cite{Zaqa1997}, the synchronization of the solar cycle 
was essentially based on some variation of the rotation rate in the 
tachocline and/or the convective layers of the Sun.
This leads to the questions: how can 
the solar dynamo be synchronized by such a weak variation 
of $\Omega$? Both from our above estimation of the tidal effect on
the velocity, as well as from the helioseismological
measurements \citep{Howe2009}, we can infer that the variation 
of the $\Omega$ effect is certainly not larger than 1 per cent, 
and very likely 
much smaller, since a significant portion of the 
$\Omega$ variation can also be attributed 
to the back-reaction of the self-excited field on the 
flow \citep{Proctor2007}.
If such a minor change is not sufficient to entrain the 
entire solar dynamo, can we perhaps take resort 
to the idea of  \cite{Abreu2012} 
who emphasized that the maximum field strength of flux tubes 
(which 
can be stored prior to eruption) is very sensitive to small 
perturbations by gravitational, tidal and, as we add here, centrifugal 
forces due to changes of $\Omega$?
Although our preliminary efforts \citep{Stefani2018} 
to implement synchronization mechanisms
of this sort into a Babcock-Leighton type dynamo model with time delay 
\citep{Wilmotsmith2006}, either 
via a direct 11-07-years variation of the $\Omega$ effect or
a corresponding variation of the rise condition for 
flux tubes, have not been very promising so far, we do not 
exclude that greater success may result from 
future investigations.

Still another avenue for synchronization was opened 
by \cite{Weber2015,Stefani2016}, who recognized 
that the intrinsic helicity oscillations
of the current-driven, kink-type Tayler instability 
\citep{Tayler1973,Pittstayler1985,Seilmayer2012,Weber2013},
characterized by an azimuthal wavenumber $m=1$, 
can be entrained by a tide-like ($m=2$)
perturbation, without (or barely) changing the 
energy content of the $m=1$ mode.
This idea of an ``energy-efficient'' mechanism of 
helicity synchronization was then first incorporated into a 
simple ODE solar dynamo model 
\citep{Stefani2016,Stefani2017,Stefani2018}, 
and later into a 1D PDE system with the co-latitude as 
the only spatial variable \citep{Stefani2019}. 
From the 11.07-years tidal entrainment 
of the helicity, and the $\alpha$-effect associated with it, 
these models produced dipolar fields with an oscillatory 
22.14-years Hale cycle, 
although in some parameter regions quadrupolar and 
hemispherical fields were observed, too. 
For the somewhat academic case of a purely 11.07-years 
periodic $\alpha$-effect we proved the existence of a 
massively nonlinear dynamo of the Tayler-Spruit type 
\citep{Spruit2002,Ruediger2020}, whereas for a 
more realistic {\it hybrid} dynamo,  
comprising both an externally ``shaken'' and 
an internally ``stirred'' $\alpha$-term, 
synchronization was accomplished by parametric resonance.

We would like to point out that meanwhile 
the empirical evidence for an 
11.07-years synchronization is quite impressive, though 
not accepted (or not even recognized) 
throughout the solar dynamo community.
Not only have the cycle minima from the last 1000 years 
turned out to be very close to a clocked-process 
with 11.07-years periodicity  \citep{Stefani2019}, but various 
algae growth data from a 1000-years period 
in the early Holocene have also shown a phase coherent 
cycle with basically the same period \citep{Vos2004}. 

As longer-term cycles are concerned, it was recently 
confirmed \citep{Stefani2020a} 
that the modulation period of the duration of the Schwabe cycles, 
as inferred from Schove's maxima data \citep{Schove1983}, 
is close to 200 years, a number which is consistent with previous 
results for the Suess-de Vries cycle relying on historic 
sunspot observations  
\citep{Ma2020}, $^{10}$Be and $^{14}$C data \citep{Muscheler2007}, 
and various climate related data \citep{Luedecke2015}. It was not 
least the relative 
sharpness of that Suess-de Vries cycle which had motivated 
many authors 
\citep{Jose1965,Fairbridge1987,Charvatova1997,Landscheidt1999,Abreu2012,Wolff2010,McCracken2014,Cionco2015,Scafetta2016} 
to search for a link between the solar dynamo and planetary forcings 
with correspondingly long periods.

Yet, when entering this playing field one can hardly avoid the 
question why not to consider, first and foremost, the strongest 
of all planetary influences on the Sun's motion, namely 
the 19.86-years synodic cycle of
Jupiter and Saturn. This cycle governs the orbit
of the Sun around
the barycenter of the planetary system, 
comprising vast deflections in the order of the Sun's diameter
and velocities of up to 15 m/s \citep{Sharp2013,Cionco2018}. Superposed 
on that period are 
minor wiggles stemming mainly from the orbits of Uranus and 
Neptun, which ultimately leads to a rather complicated motion with 
another 172-years periodicity, sometimes called ``Jose cycle''  
\citep{Jose1965,Charvatova1997,Landscheidt1999,Sharp2013}.
Still, it is the dominant 19.86-years cycle which has the 
capacity to produce, in concert with the 22.14-years 
Hale cycle, a beat period of $19.86 \times 22.14/(22.14-19.86)=193$\, 
years, as worked out in the above mentioned paper 
by  \cite{Wilson2013} and also by \cite{Solheim2013}. 
This 193-years period is suspiciously close to the Suess-de Vries cycle. 
Assuming an appropriate, though not yet well understood, 
coupling effect of the Sun's orbital motion into some change of the 
stratification of the tachocline, this beat period was 
numerically found to produce a 193-years modulation of the 
North-South asymmetry of the dynamo field 
\citep{Stefani2020a}. 

This sequel to \cite{Stefani2019,Stefani2020a} is structured as 
follows: in Section 2 we recapitulate our 1D solar dynamo code, 
and motivate the choice and structure of its main ingredients, 
such as the $\Omega$ effect, 
two parts of the $\alpha$-effect, and the time variation of the
loss parameter $\kappa$. The results of 
our simulations over 30 kyears are then presented in 
Section 3. We will illustrate how a weak time-variation 
of $\kappa$  produces a clear 193-years modulation of the 
North-South-asymmetry, and the Gnevyshev-Ohl effect associated with 
it.  For stronger variations of $\kappa$ we then obtain intermittent 
breakdowns of the solar cycle which are indeed reminiscent of 
grand minima, and clusters thereof. 
Since those breakdowns occur already in the absence of any noise,
we argue here for the onset of an intermittent route to chaos. 
One of those long runs will
be utilized to produce the numerical time series
of breakdowns as shown in Figure 1.
We will conclude with a number of suggestions for future work,
including an urgent call for a better physical and numerical 
modelling of the two main synchronization mechanisms, 
which in the present paper can only be implemented in a 
parameterized form.

\section{Numerical model}

Inspired by early work 
of \cite{Parker1955,Schmalz1991,Jennings1991,Roald1997,Kuzanyan1997},
we will use here the same system of PDSs as in 
\cite{Stefani2019,Stefani2020a}, with the solar co-latitude 
$\theta$ as the only spatial coordinate.
While appropriately constructed ODE systems have been 
surprisingly successful in describing different types 
of dynamo modulations \citep{Knobloch1998}, and even supermodulation
\citep{Weiss2016}, 
we consider a 1D PDE model a reasonable intermediate step 
towards  a 2D model (in $r$ and $\theta$) which is 
presently under development, but which might
become numerically costly when aiming at
very long dynamo runs over 30 kyears, say.

With the axisymmetric
magnetic field split into a
poloidal component ${\bf{B}}_P=\nabla \times (A {\bf{e}}_{\phi})$
and a toroidal component ${\bf{B}}_T=B {\bf{e}}_{\phi}$, 
the 1D PDE system reads
\begin{eqnarray} 
  \frac{{\partial} B(\theta,t)}{{\partial} t} &=& \omega(\theta,t) \frac{\partial A(\theta,t)}{\partial \theta} 
  + \frac{\partial^2 B(\theta,t)}{\partial \theta^2} -\kappa(t) B^3(\theta,t) \\
    \frac{{\partial} A(\theta,t) }{{\partial} t} &=& \alpha(\theta,t) B(\theta,t) 
    + \frac{\partial^2 A(\theta,t)}{\partial \theta^2} , 
    \label{system_tayler}
   \end{eqnarray}
where $A(\theta,t)$ represents the vector potential of the 
poloidal field at co-latitude $\theta$ (running between 0 and $\pi$) 
and time $t$, and $B(\theta,t)$ the corresponding toroidal field. 
The two sources of dynamo action are
the helical turbulence parameter $\alpha$ and
the radial derivative $\omega=\sin(\theta) d (\Omega r)/dr$
of the rotational profile.
Here, $\alpha$ and $\omega$ 
denote the non-dimensionalized
versions of the dimensional quantities $\alpha_{\rm dim}$ and 
$\omega_{\rm dim}$,
according to $\alpha=\alpha_{\rm dim} R/\eta$  and
$\omega=\omega_{\rm dim} R^2/\eta$, where $R$ is the radius of the
considered dynamo region and $\eta$ the magnetic diffusivity. 
The time is non-dimensionalized by
the diffusion time, i.e. $t=t_{\rm dim} \eta/R^2$.
The boundary conditions at the North and South pole 
are $A(0,t)=A(\pi,t)=B(0,t)=B(\pi,t)=0$.
The PDE system is solved by a finite-difference scheme
using the Adams-Bashforth method.
The initial conditions are
$A(\theta,0)=0$ and $B(\theta,0)=s \sin(\theta)+ u \sin(2 \theta)$,  
with the chosen pre-factors $s=-1$ and $u-0.001$ 
denoting symmetric and asymmetric 
components of the toroidal field\footnote{Note that in \cite{Stefani2019}
the value for $A$ was erroneously indicated as non-zero.}.

We employ the typical solar $\theta$-dependence of the $\omega$-effect 
\citep{Charbonneau2010}
in the form 
\begin{eqnarray}
\omega(\theta)&=&\omega_0 (1-0.939-0.136 \cos^2(\theta)-0.1457 \cos^4(\theta) )\sin(\theta)
\end{eqnarray}
with a plausible, but still moderate, value 
$\omega_0=10000$. The helical source term $\alpha$ comprises, first, a non-periodic 
part 
\begin{eqnarray}
\alpha^c(\theta,t)&=&\alpha^c_0(1+\xi(t))  \sin(2 \theta)/{(1+q^c_{\alpha} B^2(\theta,t))}
\end{eqnarray}
with a constant $\alpha^c_0$ and a noise term $\xi(t)$
(which is defined by the correlator $\langle \xi(t) \xi(t+t_1) \rangle = D^2 (1-|t_1|/t_{\rm corr})
\Theta(1-|t_1|/t_{\rm corr})$), and second,
a periodic part
\begin{eqnarray}
\alpha^p(\theta,t)&=&\alpha^p_0 \sin(2 \pi t/t_{11.07}) 
 \frac{B^2(\theta,t)}{(1+q^p_{\alpha} 
B^4(\theta,t))} S(\theta)\; \mbox{for $55^{\circ}<\theta<125^{\circ}$} \nonumber \\
&=&0 \; \mbox{elsewhere} \; ,
\end{eqnarray}
where the $B$-dependent term has the 
typical resonance-type 
structure $\sim B^2/(1+q^p_{\alpha} B^4)$.
The expression $t_{11.07}$ denotes the dimensionless
counterpart of the 11.07-year tidal forcing period.
With our special choice of the diffusion time 
$R^2/\eta=110.7$\,years, this amounts to $t_{11.07}=0.1$.
Note that the latitudinal dependence 
\begin{eqnarray} 
S(\theta)&=&
{\rm{sgn}}(90^{\circ}-\theta) \tanh^2\left(  \frac{\theta/180^{\circ}-0.5}{0.2}  \right)
\end{eqnarray}
comprises the same smoothing term (although more 
conveniently written here) 
as in \cite{Stefani2019},
which avoids a steep jump of $\alpha$ 
at the equator.

The  term $\kappa(t) B^3(\theta,t)$ in Eq. (1), 
as originally introduced by \cite{Jones1983,Jennings1991}, 
has been included to account 
for field losses owing to magnetic buoyancy, on the 
assumption that the escape velocity is 
proportional to $B^2$.  While we 
openly admit that the spin-orbit coupling 
of the angular momentum of the Sun around the barycenter
into some dynamo 
relevant parameters remains an open question (for ideas, 
see \cite{Zaqa1997,Palus2000,Wilson2008,Sharp2013}) 
we employ in the following a time-variation of the parameter
$\kappa(t)$  proportional to the time series of the angular momentum. 
Since $\kappa(t)$  is related to 
the very sensitive adiabaticity in the tachocline \citep{Abreu2012}, 
which could be easily influenced by slight changes in the
internal rotation profile, its modification by some sort
of spin-orbit coupling seems not completely 
unrealistic.

\begin{figure}[t]
  \centering
  \includegraphics[width=0.99\textwidth]{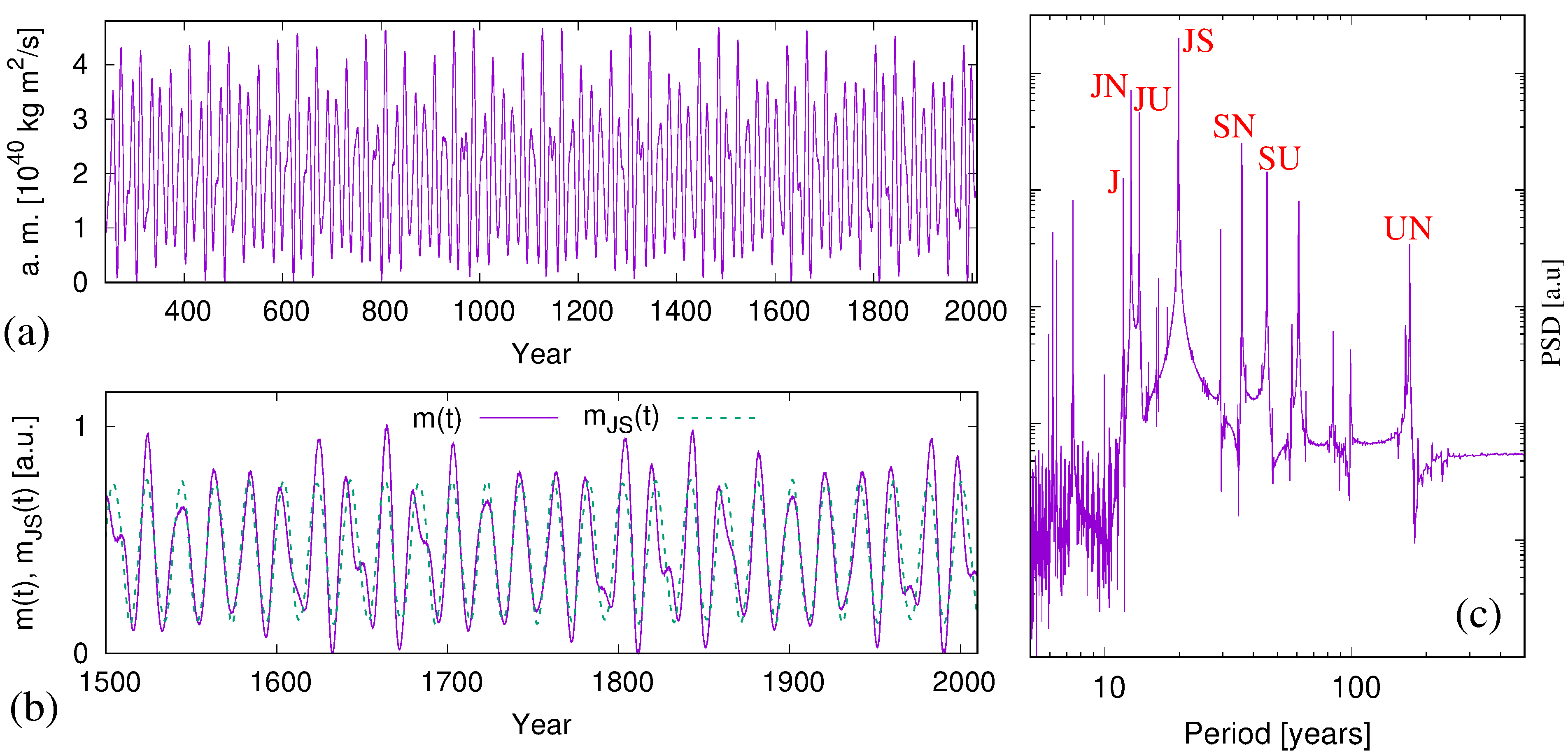}
  \caption{(a) Time series of the orbital angular momentum (a.m.) 
  of the Sun around the barycenter of the solar system 
  during the interval A.D. 240-2010, based on the DE431 ephemerides  
  \citep{Folkner2014}. 
  (b) Zoom of (a) for the interval  A.D. 1500-2010, normalized to 1,
  giving the modulation function $m(t)$ (violet full line). The corresponding function
  $m_{\rm JS}$ (green dashed) is a theoretical line that would result from 
  the exclusive action of Jupiter and Saturn. 
  (c) PSD of the angular momentum for the long interval 13199 B.C.-A.D. 17000, with 
  some individual peaks attributed to planetary synodes (cf. \cite{Scafetta2016}): 
  JN: Jupiter-Neptune (12.78 years), JU: Jupiter-Uranus (13.95 years), 
  JS: Jupiter-Saturn (19.86 years), 
  SN: Saturn-Neptune (35.87 years), SU: Saturn-Uranus (45.36 years),
  UN: Uranus-Neptune (171.39 years). J indicates the 11.86 years period 
  of Jupiter.
  }
  \label{Fig:angular}
\end{figure}

For the computation of the Sun's orbital angular momentum we utilized
the DE431 ephemerides  \citep{Folkner2014} in the 
time interval between 13199 B.C.-A.D. 17000, of 
which a $\approx$1800-years segment
is visualized in Figure \ref{Fig:angular}(a).
This function is dominated by the 19.86-years synodes of Jupiter and Saturn,
to which further contributions, mainly from Uranus and 
Neptun, are added (see the PSD in Figure \ref{Fig:angular}(c)). 
Further below, we will use the normalized version $m(t)$ of this 
angular momentum curve 
for parametrizing the time-variation of $\kappa(t)$. 
For the sake of comparison, 
we will also assess the results for a modified variant $m_{\rm JS}(t)$
which relies exclusively on the 19.86-years periodic 
motion of Jupiter and Saturn.

\section{Results}

In this section, we present and discuss numerical results for a 
sequence of parameters similar to those utilized in \cite{Stefani2020a}.
The main difference is the much longer simulation time of 30 kyears, 
which is actually the interval for which the
orbital angular momentum of the Sun was available to us 
\citep{Folkner2014}.
Such longer simulations will allow for a more systematic 
study of the typical breakdowns of the 193-years modulated wave,
and some preliminary comparisons with the sequence of Bond events.
Another difference to  \cite{Stefani2020a} is that here we
focus chiefly on the noise-free
case in order to evidence the intermittent route to deterministic
chaos.  A few results with noise included will nevertheless be presented 
at the end of the section and in the Appendix.

\subsection{All planets, no noise}

\begin{figure}[!ht]
\includegraphics[width=0.98\textwidth]{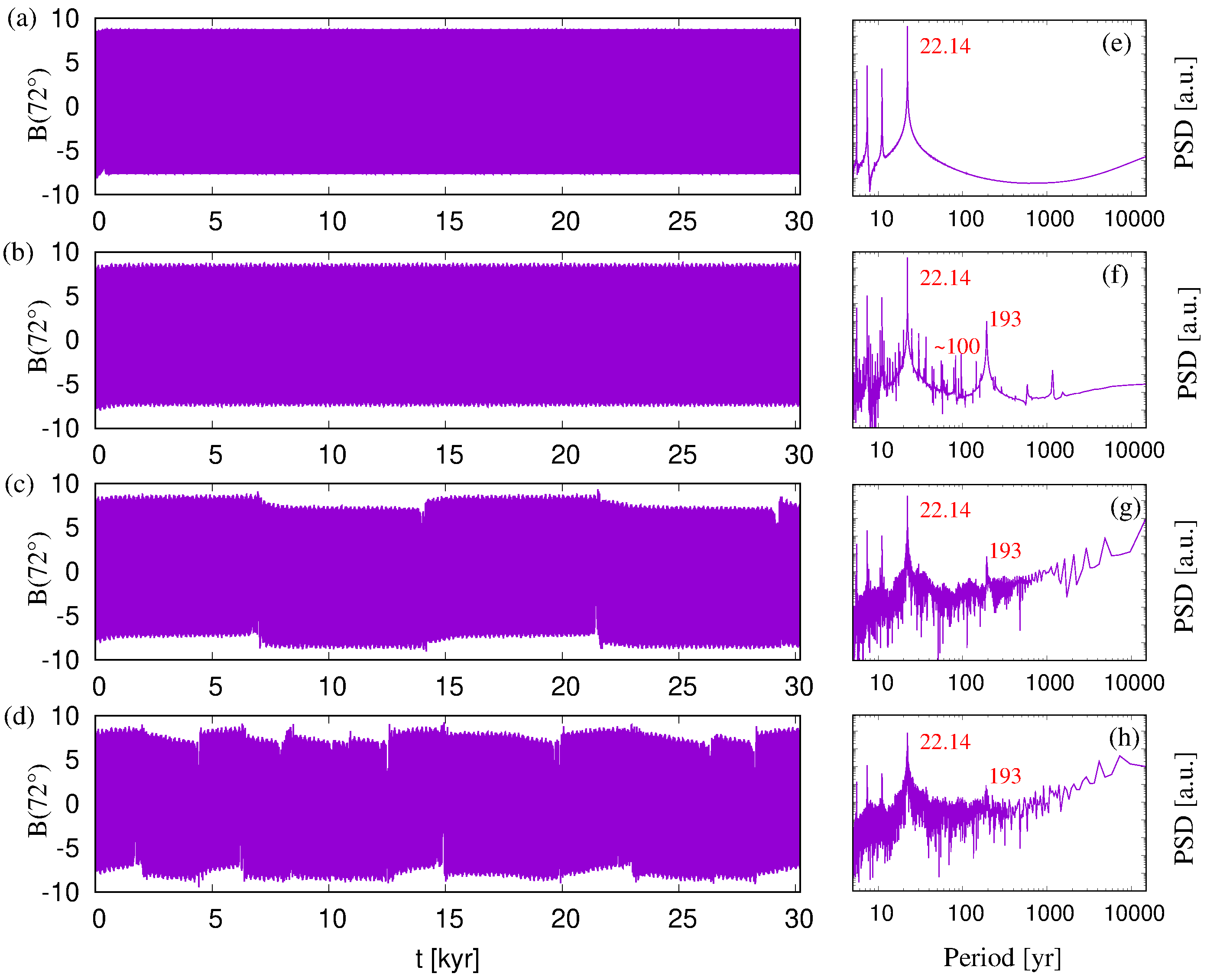}
\caption{(a-d) Time evolution of $B(72^{\circ},t)$, and its Lomb-Scargle 
PSD  (e-h), with the common parameters
  $\omega_0=10000$, $\alpha^c_0=15$, $\alpha^p_0=50$
  $q^p_{\alpha}=0.2$, $q^c_{\alpha}=0.8$, $D=0$.
  The time variation of the parameter $\kappa(t)$ 
  increases from top to bottom.
  (a,e): $\kappa(t)=0.6$,  a tidally synchronized
  dynamo producing a dipole with 22.14-year period.
  Note already the slight positive-negative asymmetry of $B(72^{\circ},t)$.
  (b,f): $\kappa(t)=0.6+0.3 m(t)$, as (a,e), but with
  a modulation of $\kappa$ with an angular momentum function $m(t)$
  according to Figure \ref{Fig:angular}(b).
  As seen in the spectrum (f), this dipole solution contains a beat period of 193 years, 
  which also appears in (b) as a minor wiggle of $B(72^{\circ},t)$. Note the 
  appearance of Gleissberg-type 
  cycles around 100 years in (f).
  (c,g): $\kappa(t)=0.6+0.32 m(t)$,  as (b,f), but with a 
  slightly increased variation of 
  $\kappa$. Note the appearance in (c) of four sudden events, where 
  the positive-negative asymmetry of $B(72^{\circ},t)$
  changes sign. The spectrum (g) has become noisy.
  (d,h): $\kappa(t)=0.6+0.4 m(t)$, as (c,g), but with 
  increased variation of $\kappa$, producing significantly more 
  breakdowns. }
\label{Fig:fiveexamples}
\end{figure}

Figure \ref{Fig:fiveexamples} summarizes 
four different dynamo simulations carried out over an interval of 
30.2\,kyears. 
The time series of  $B(\Theta=72^{\circ},t)$ are illustrated in the 
left panels (a-d), whose ``patchy'' appearances  
simply result from the large number ($\sim 1350$) of Hale cycles 
involved (more details will be shown further below).
The right panels (e-h) show the associated power spectral densities (PSD)
resulting from Lomb-Scargle analyses of the respective curves in (a-d).

The fixed parameters
$\omega_0=10000$, $\alpha^c_0=15$, $\alpha^p_0=50$ 
$q^p_{\alpha}=0.2$, $q^c_{\alpha}=0.8$, $D=0$ are chosen
similar to those of previous studies 
\citep{Stefani2019,Stefani2020a}, but note the complete 
absence of noise in the present runs. Moreover, here we set out with a 
sufficiently large value of $\alpha^p_0$ so that the dynamo is already 
synchronized to 22.14 years; the 
parametric resonance phenomenon behind the 
frequency synchronization, when going over 
from  $\alpha^p_0=0$ to some finite value, had been discussed 
in detail in  \cite{Stefani2019,Stefani2020a} and will not 
be re-iterated here. Hence, the only parameter to be varied from 
top to bottom of Figure \ref{Fig:fiveexamples}
is the loss parameter $\kappa$. 

We start in Figure \ref{Fig:fiveexamples}(a,e) with a 
time-independent value
$\kappa(t)=0.6$, which yields a very clean Hale cycle with 
a period 22.14 years. Already in panel (a) we observe a slight 
asymmetry between positive values (reaching $\approx 9$) and negative 
values (reaching $\approx -8$) that can be attributed to 
the presence of a mixed mode, in which a weak quadrupolar 
field component part is added to the dominant dipolar one. 
This effect is related to the Gnevyshev-Ohl rule 
\citep{Gnevyshev1948}.
Apart from the two minor peaks at the doubled and tripled Hale 
frequency (which naturally result from the nonlinear terms in the
PDE system) the spectrum in (e) is quite smooth and featureless.

This is changing in Figure \ref{Fig:fiveexamples}(b,f) when the loss 
parameter is equipped with a
time-variation according to $\kappa(t)=0.6+0.3 m(t)$, where $m(t)$ is 
the normalized orbital angular momentum function from Figure \ref{Fig:angular}(b). 
The most prominent 
feature that appears in the PSD (f) is the 
strong peak at 193 years. In panel (b) this peak manifests
itself in form of minor wiggles of the maxima and minima, which indeed
correspond to a modulation of the positive-negative
asymmetry. We note in passing the occurrence of some 
peaks at Gleissberg-type periods (around 96 years and 64 years), and 
two side bands at around  20 and 25 years which are reminiscent 
of the Wilson gap \citep{Hathaway2010}. These two features were
discussed in more detail in \cite{Stefani2020a}, and will not 
play a particular role in the following.

When increasing the variation of the loss parameter
just a little further to $\kappa(t)=0.6+0.32 m(t)$, 
we observe in panel (c) four sudden jumps
of the positive-negative asymmetry. While the main 
peaks at 22.14\,years and 193\,years survive, the rest 
of the spectrum (g) becomes noisy, an effect
of spectral leakage due to the segmentation of the entire time domain 
into 5 parts. While
in this case one might still speculate about a regular
occurrence of these breakdowns (with a waiting time of appr. 7 kyears),
any alleged regularity is clearly 
lost in our last example 
(d,h), corresponding to $\kappa(t)=0.6+0.4 m(t)$. 
Here we observe approximately 11
breakdowns without any clear periodicity. This 
insinuates an intermittent route to chaos, although 
we leave the detailed 
analysis of this  transition, for our non-trivial case of 
double synchronized system, to future work.

\begin{figure}[!ht]
\includegraphics[width=0.98\textwidth]{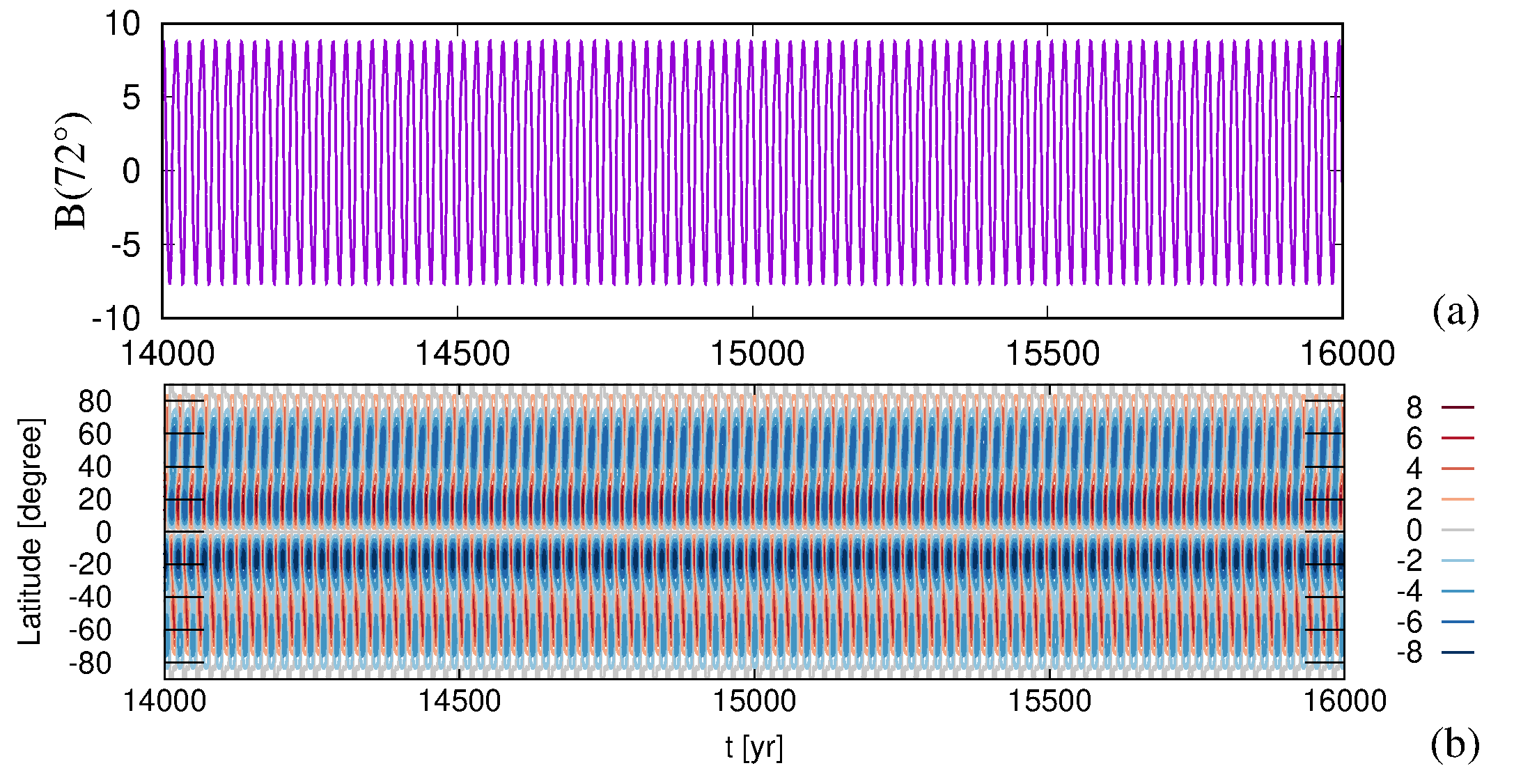}
\caption{(a) Details of Figure \ref{Fig:fiveexamples}(a) for the central interval 
between 14000 and 16000 years, with parameters
$\omega_0=10000$, $\alpha^c_0=15$, $\alpha^p_0=50$
  $q^p_{\alpha}=0.2$, $q^c_{\alpha}=0.8$, $D=0$, $\kappa=0.6$.
 (b) Contour-plot of $B(\theta,t)$ for the same parameters.
 Note that in (b) the ordinate axis represents not the 
 colatitude $\theta$ but the normal solar latitude  $90^{\circ}-\theta$.}
 \label{Fig:kappavar0}
\end{figure}

For three of the examples from Figure \ref{Fig:fiveexamples}, some 
details are discussed in the following.
Figure  \ref{Fig:kappavar0} shows the central interval between 
14 and 16\,kyears from
Figure  \ref{Fig:fiveexamples}(a), both for $B(72^{\circ},t)$ (a) and
for the entire field $B(\theta,t)$ as shown here as a contour-plot (b).
In panel (a) we observe again the positive-negative asymmetry 
(the value varies between -8 and +9), which translates into a 
North-South asymmetry as visible in (b) by the
color asymmetry between Northern and Southern regions. Evidently
this asymmetry is also connected with a Gnevyshev-Ohl rule.

\begin{figure}[!ht]
\includegraphics[width=0.98\textwidth]{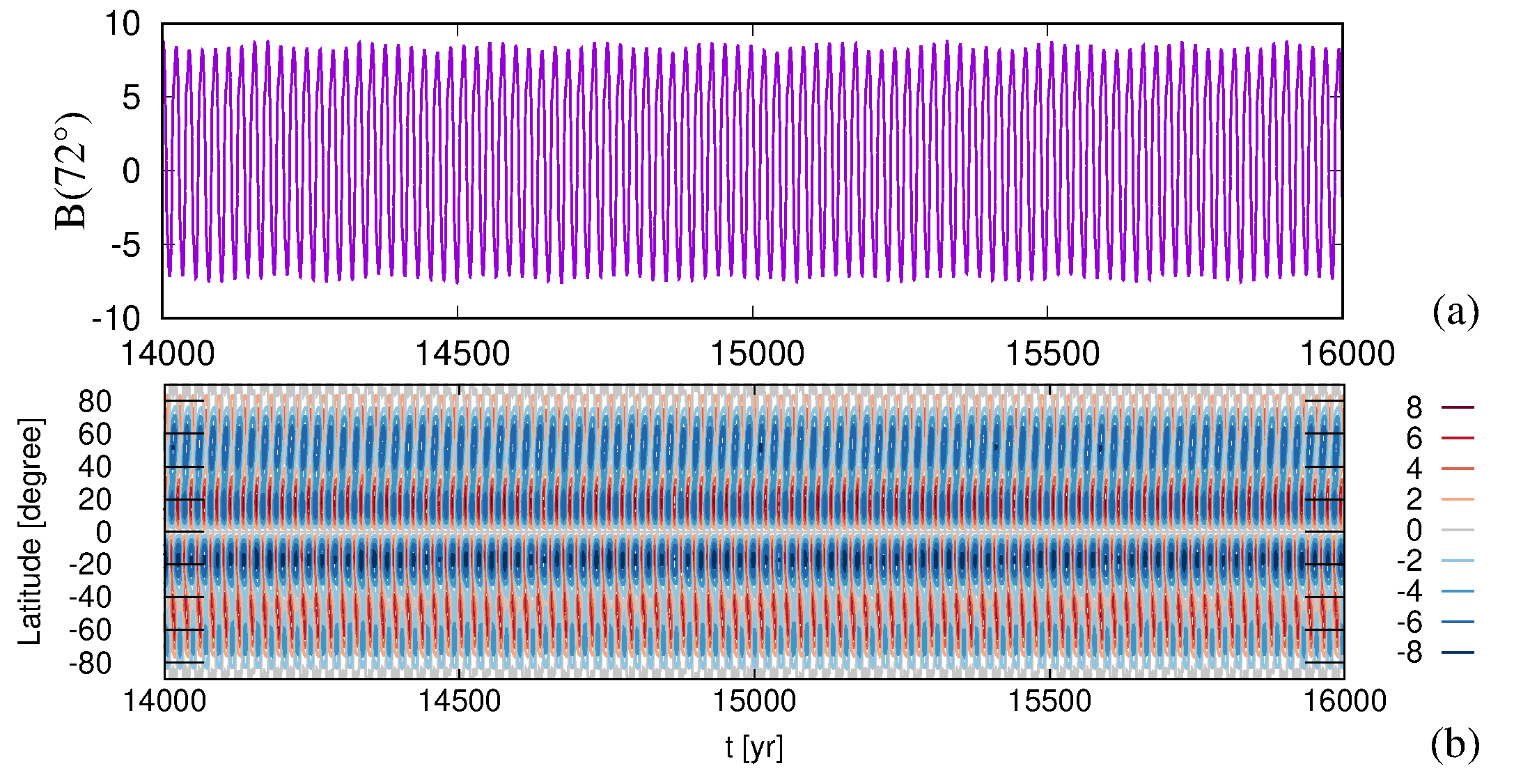}
\caption{Same as Figure \ref{Fig:kappavar0}, but for Figure \ref{Fig:fiveexamples}(b)
with $\kappa=0.6+0.3 m(t)$.}
\label{Fig:kappavar03}
\end{figure}

\begin{figure}[!ht]
\includegraphics[width=0.98\textwidth]{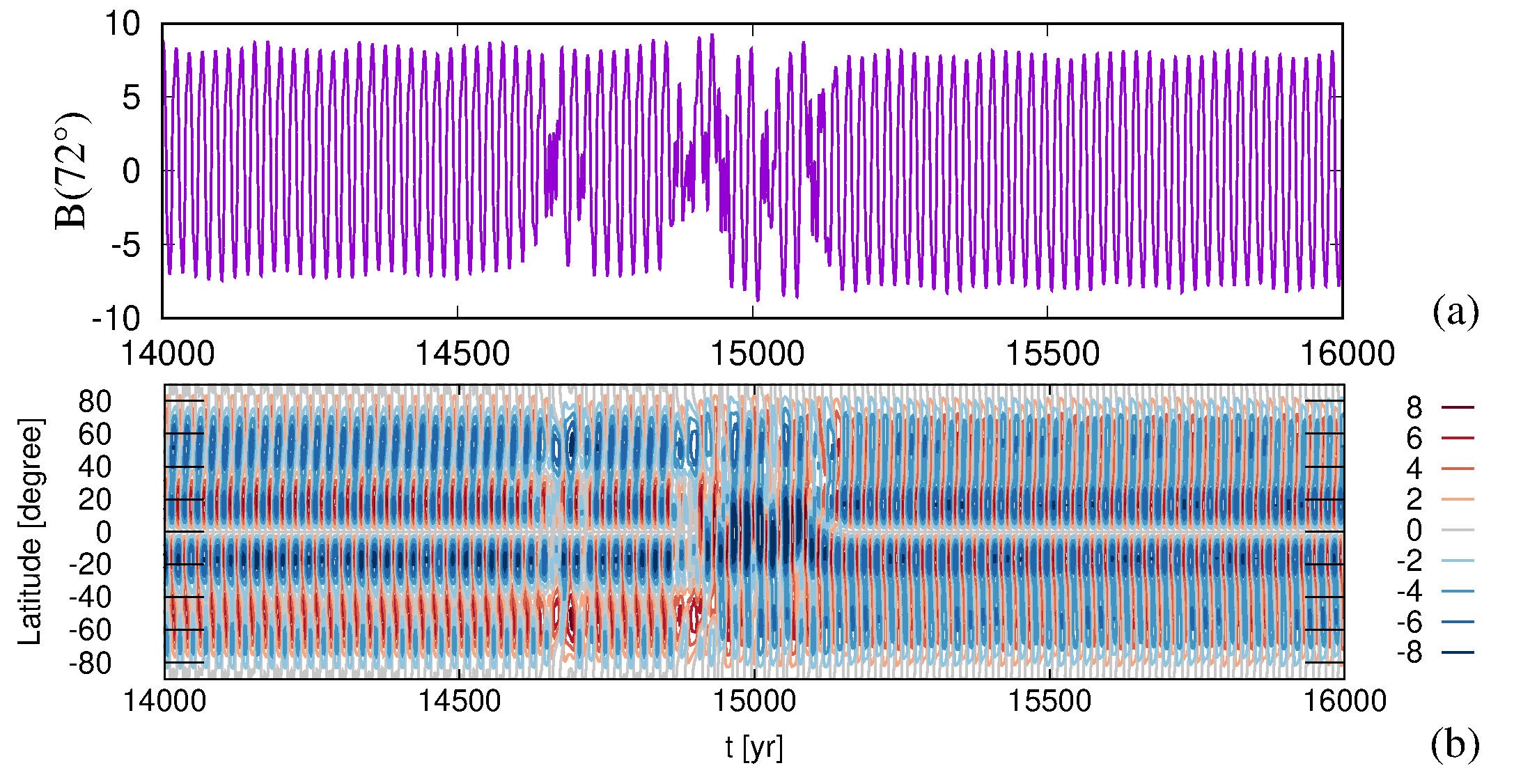}
\caption{Same as Figure \ref{Fig:kappavar0}, but for Figure \ref{Fig:fiveexamples}(d)
with $\kappa=0.6+0.4 m(t)$.}
\label{Fig:kappavar04}
\end{figure}

\begin{figure}[!ht]
\centering
\includegraphics[width=0.8\textwidth]{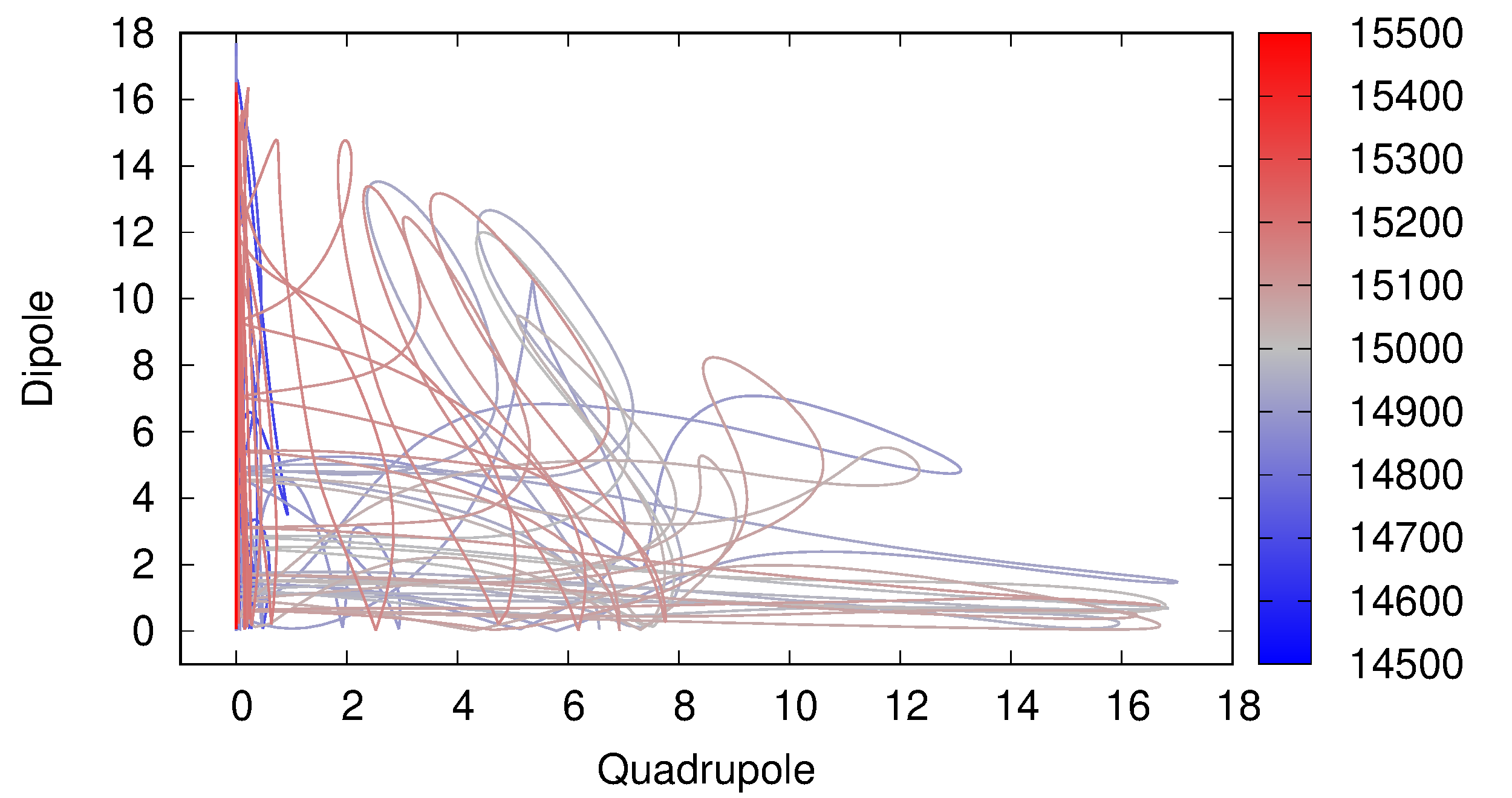}
\caption{Trajectory of the solution from 
Figure \ref{Fig:kappavar04} for the shortened 
interval between 14500-15500\,years. The abscissa shows the quadrupolar component 
defined here as $|B(72^{\circ})+B(108^{\circ})|$, the corresponding dipolar
component $|B(72^{\circ})-B(108^{\circ})|$ is shown on the ordinate axis.}
\label{Fig:dipole-quadrupole}
\end{figure}

A clear 193-years modulation of the positive-negative asymmetry, and the 
Gnevyshev-Ohl effect related to it,
appears in Figure \ref{Fig:kappavar03} 
which illustrates some details of Figure \ref{Fig:fiveexamples}(b).
For the central interval of Figure \ref{Fig:fiveexamples}(d), with 
$\kappa=0.6+0.4 m(t)$, Figure \ref{Fig:kappavar04}
shows one typical breakdown of the modulated wave. The resulting 
disordered state, lasting appr. 500\,years, resembles indeed
a cluster of grand minima where the dynamo is not 
switched off \citep{Beer1998}, but just
in another state. Quite interesting here is the 
appearance of hemispherical fields (around 14900) and quadrupole 
fields (around 15000), which are reminiscent of sunspot 
observations within or shortly after the Maunder minimum  
\citep{Sokoloff1994,Arlt2009}.
The corresponding trajectory in the dipole-quadrupole space,
as shown in Figure \ref{Fig:dipole-quadrupole}, resembles strongly 
the corresponding behaviour in Figure 5a of \cite{Knobloch1998}
and Figure 4 of \cite{Weiss2016}.
Meanwhile, a similar supermodulation effect has a also been found in 
a 3D simulation of dynamo action in rotating anelastic convection 
\citep{Raynaud2016}.

\subsection{Only Jupiter and Saturn, no noise}

We now consider a modification of the 
angular momentum that enters the time-variation of the
loss parameter $\kappa(t)$. While in the previous subsection the full
(normalized) angular momentum curve $m(t)$ 
was used (violet full line in Figure \ref{Fig:angular}(b)), 
we consider now 
the idealized curve $m_{\rm JS}(t)$ as it would result from 
exclusively taking into account the orbital motion of 
Jupiter and Saturn (green dashed curve in Figure 
\ref{Fig:angular}(b)).

\begin{figure}[!ht]
\includegraphics[width=0.98\textwidth]{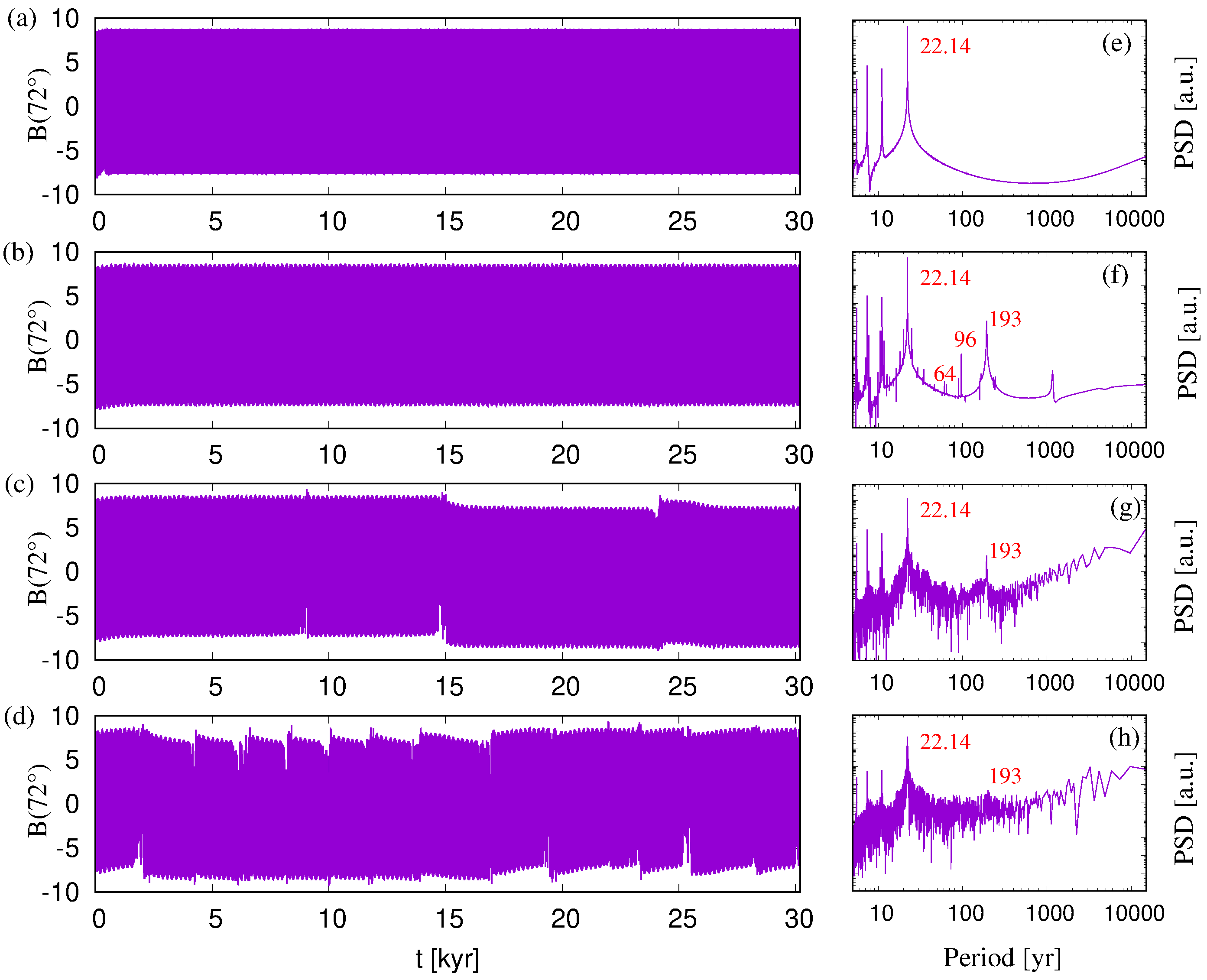}
\caption{Same as Figure \ref{Fig:fiveexamples}  but with the 
simpler angular momentum variation $m_{\rm JS}(t)$ 
(green dashed line in 
Figure \ref{Fig:angular}(b)) 
 which is restricted to the 19.86-years periodic 
 part resulting from the orbital motion of 
 Jupiter and Saturn only. }
\label{Fig:simple}
\end{figure}

\begin{figure}[t]
  \centering
  \includegraphics[width=0.9\textwidth]{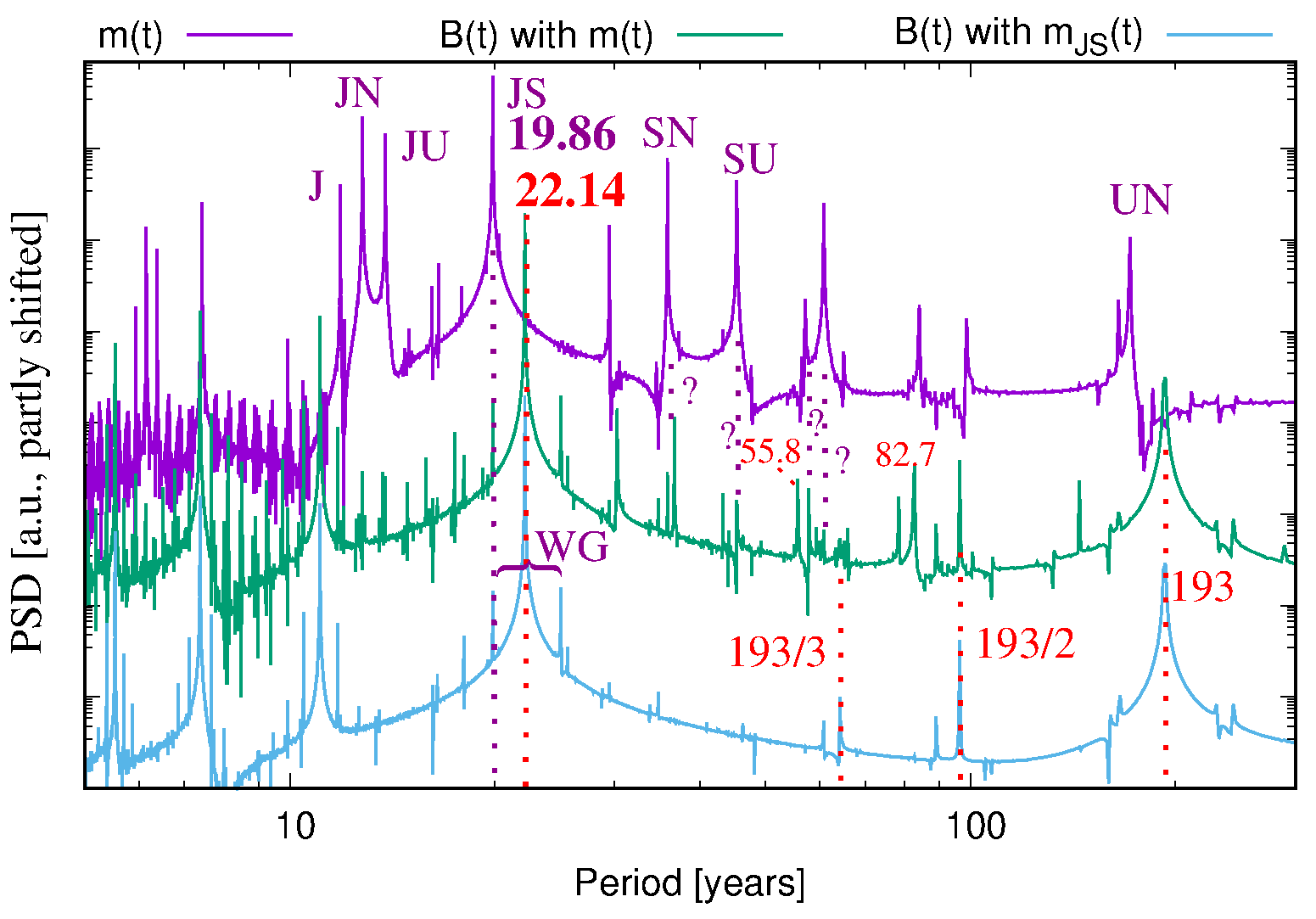}
  \caption{Comparison of the Lomb-Scargle PSDs for the 
  angular momentum
  function m(t) (violet), for $B(72^{\circ},t)$ (green) resulting 
  from $\kappa(t)=0.6+0.3 m(t)$ with the full $m(t)$, and
  for $B(72^{\circ},t)$ (blue) resulting 
  from $\kappa(t)=0.6+0.2 m_{\rm JS}(t)$ with the reduced $m_{\rm JS}(t)$. 
  The individual peaks of the 
  PSD for $m(t)$ are the same as in Figure 
  \ref{Fig:angular}(c). The Suess-de Vries period of 193
  years, as well Gleissberg-type periods 193/2 and 193/3 years 
  emerge already when using only $m_{\rm JS}(t)$.
  Some more peaks, which appear when the full  $m(t)$
  is utilized, can be attributed to corresponding peaks 
  in $m(t)$ (see the question marks). 
  However, there are two additional peaks at 55.8 years and 82.7
  years which represent beat periods between
  the 11.07-years Schwabe cycle and the  
  Jupiter-Uranus  synode  (13.81 years) and the 
  Jupiter-Neptune synode (12.78 years), respectively.
  WG denotes the Wilson gap 
  between 19.86 and 25 years.}
  \label{Fig:period_vergleich}
\end{figure}

Figure \ref{Fig:simple} shows the corresponding numerical 
results for otherwise 
the same parameters as used previously 
in Figure \ref{Fig:fiveexamples}. Superficially, 
the results are very similar, with the most significant differences 
showing up for the range of Gleissberg-type periods. 
With the simplified $m_{\rm JS}(t)$curve, we observe now
clean peaks at one half (96.5 years) and one third (64.3 years) 
of the 193-years beat period, whereas in Figure 
\ref{Fig:fiveexamples} those peaks were more 
complicated.  In Figure \ref{Fig:period_vergleich}
we summarize our present understanding regarding the origin 
of the different peaks. The PSD for $m(t)$ (violet) is a reproduction 
from Figure \ref{Fig:angular}(c). It is dominated by
the Jupiter-Saturn-peak at 19.86 years, and some other peaks, including the
Jupiter-Neptune synode (12.78 years) and the 
Jupiter-Uranus  synode  (13.81  years). 
Both for the field resulting
from $m(t)$ and from $m_{\rm JS}(t)$, the two dominant peaks 
 are the Hale period
(22.14 years) and the Suess-de Vries period (193 years).
While the (blue) field curve for $m_{\rm JS}(t)$ contains 
basically only one half (96.5 years) and one third
(64.3 years) of this Suess-de Vries period, the (green) 
 curve for the full
 $m(t)$ contains additional peaks in the Gleissberg-region, 
 comprising
 in particular the peaks at 55.8 years and 82.7
  years which are beat periods between
  the 11.07-years Schwabe cycle and the  
  Jupiter-Uranus  synode  (13.81  years) and 
  Jupiter-Neptune synode (12.78 years), respectively. 
 Moreover, a few additional peaks, indicated by question marks, 
 seem to be related, e.g.,  
 to the Saturn-Neptun (35.87 years) and the Saturn-Uranus (44.36 years)
  synodes.

\subsection{All planets, noise included}

\begin{figure}[!ht]
\includegraphics[width=0.98\textwidth]{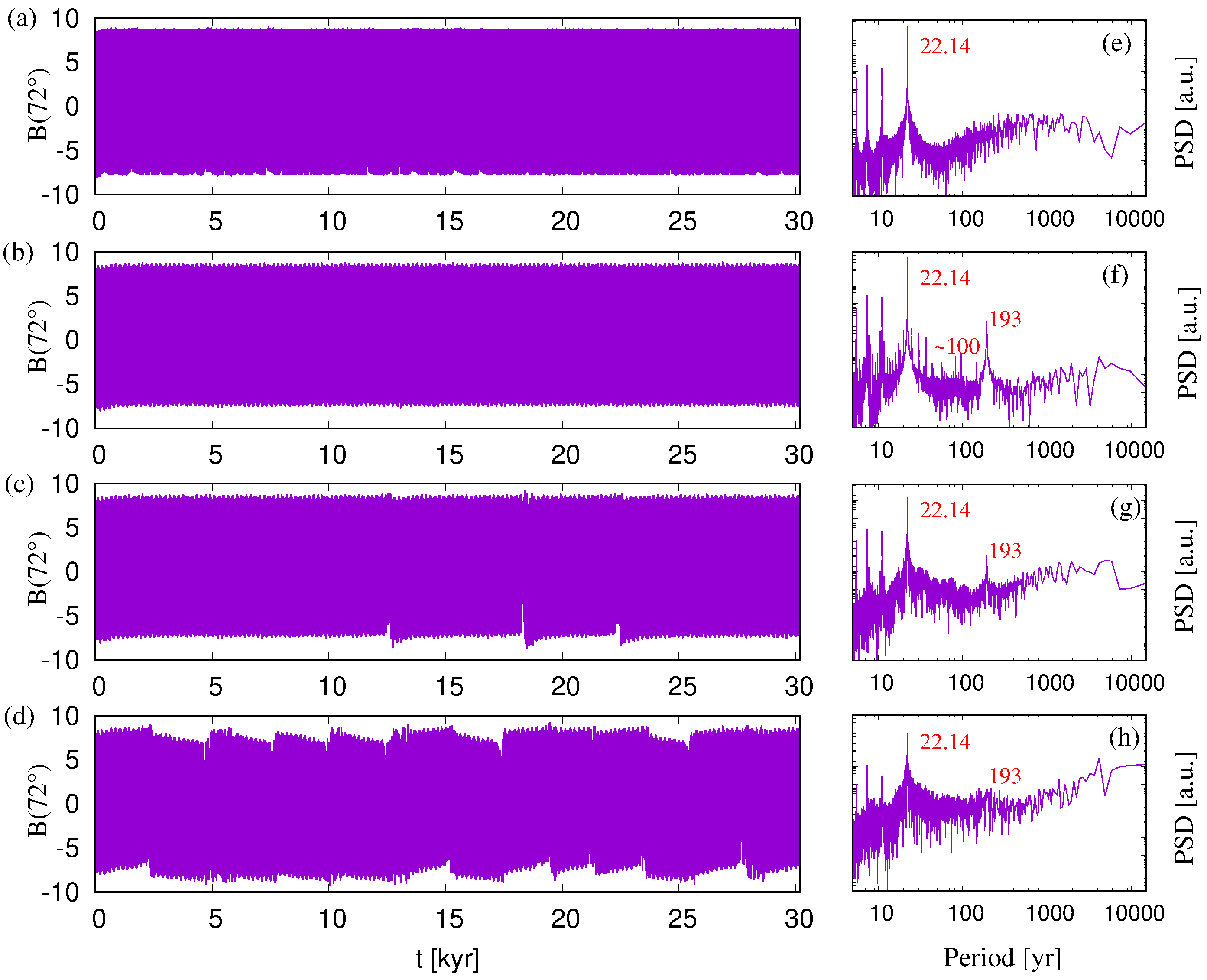}
\caption{Same as Figure \ref{Fig:fiveexamples}  but with noise intensity 
$D=0.05$. Even without any $\kappa$ variation, the spectrum (e) is 
already noisy, while the positive-negative asymmetry in (a) is 
very similar to that in \ref{Fig:fiveexamples}(a). Apart from that, 
the overall structure, 
including the critical value for the transition to chaos,
is very similar as for $D=0$.}
\label{Fig:noise}
\end{figure}

In Figure \ref{Fig:noise} we assess the role of noise.
The parameters are identical to those in Figure \ref{Fig:fiveexamples},
except that we use now a finite noise level $D=0.05$.
Not surprisingly, even without any $\kappa$ variation (a), 
the spectrum (e) is already noisy, while
the positive-negative asymmetry in (a) is very similar to 
Figure \ref{Fig:fiveexamples}(a). Apart from that, the overall structure 
turns out to be quite comparable to  Figure \ref{Fig:fiveexamples}.

\begin{figure}[!ht]
\includegraphics[width=0.98\textwidth]{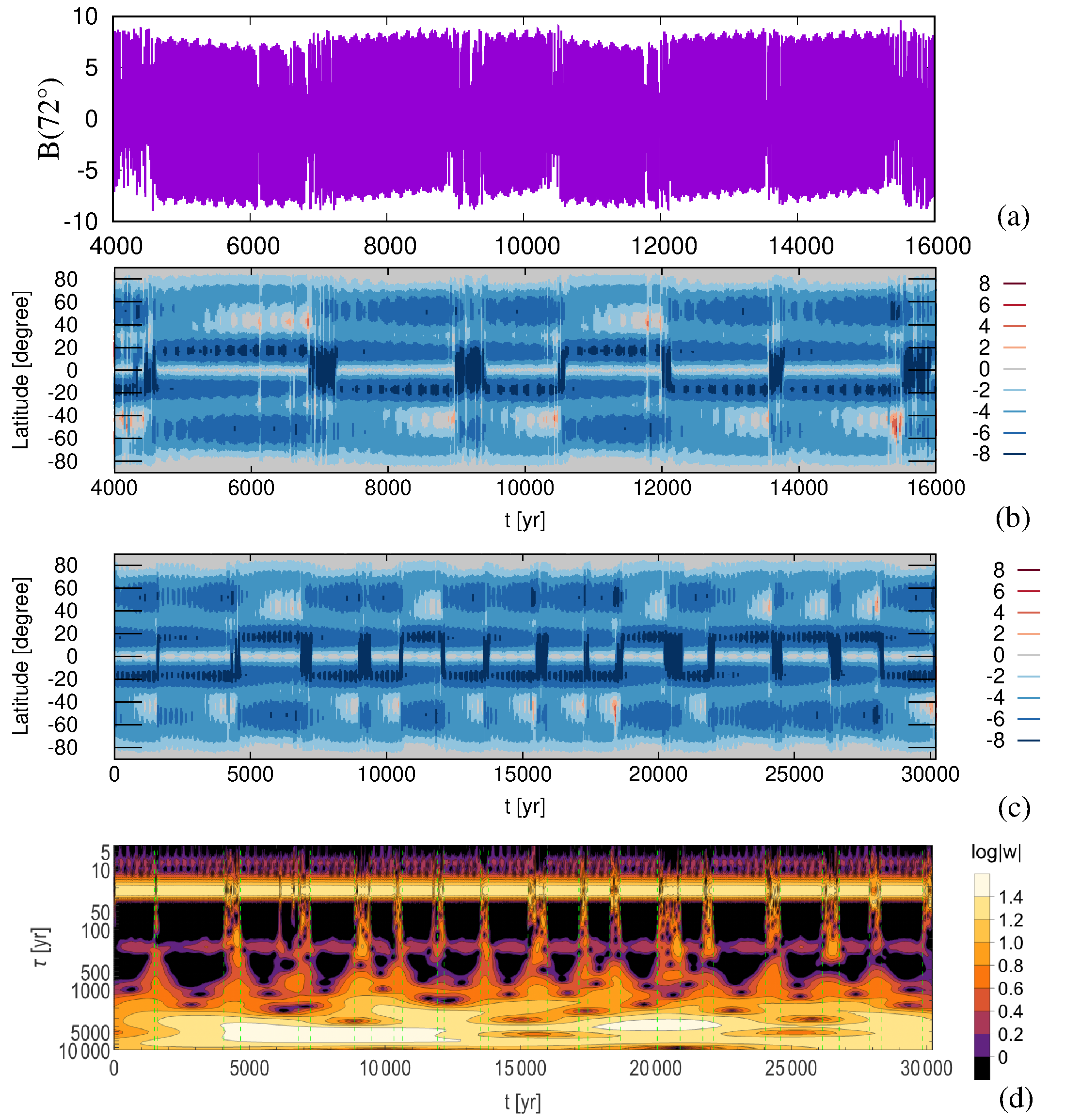}
\caption{(a) Behaviour of $B(72^{\circ},t)$ for $\omega_0=10000$, 
$\alpha^c_0=15$, $q^p_{\alpha}=0.2$, $q^c_{\alpha}=0.8$, $\alpha^p_0=50$, 
$D=0.05$ and $\kappa=0.6+0.5 m(t)$, in the time interval
4000-16000 years.
(b) Contour plot of $B(\theta,t)$ for the same parameters.
(c) Contour-plot for the full time interval
0-30200 years. (d) Wavelet diagram for the full time interval. The green dashed lines
in (d) indicate the intervals of breakdowns.}
\label{Fig:kappavar05}
\end{figure}

If we go beyond the examples of 
Figure \ref{Fig:noise} by choosing a still stronger variation 
$\kappa=0.6+0.5 m(t)$, 
we end up with Figure  \ref{Fig:kappavar05} which shows now 
a longer segment between 4-16\,kyears. 
While for such long simulations many 
details are lost in the contour-plot (b), it
highlights the fact that the breakdowns occur at instants where the 
North-South asymmetry (evidenced in particular by the
reddish parts) has acquired a certain critical threshold.
This entire behaviour is reminiscent of the supermodulation 
described by \cite{Weiss2016,Raynaud2016}.

\begin{figure}[t]
  \centering
  \includegraphics[width=0.8\textwidth]{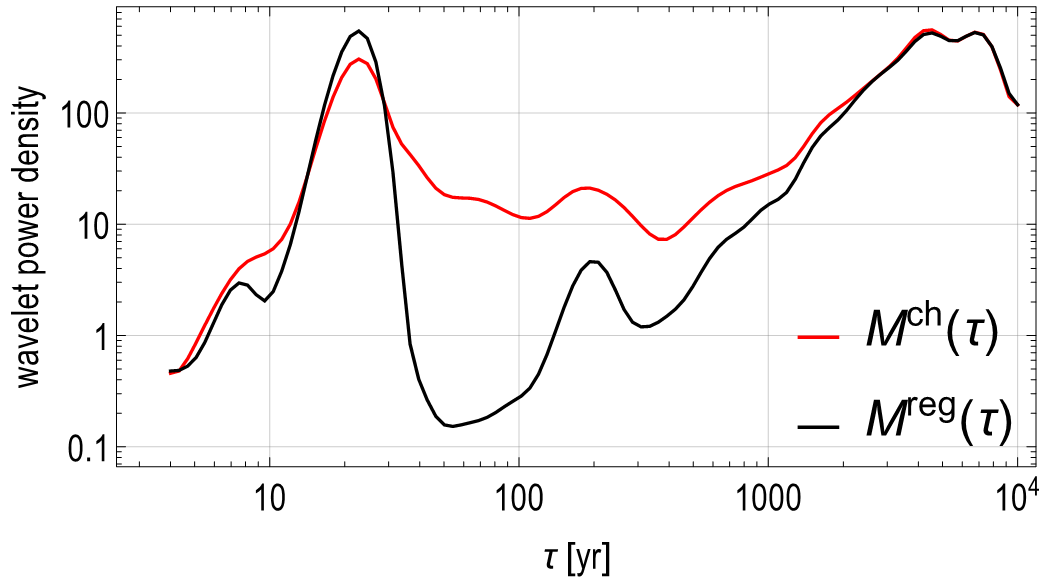}
  \caption{Comparison of the wavelet power density for the 
   ``chaotic'' intervals (separated by the green dashed lines 
   in Figure \ref{Fig:kappavar05}(d)) with that for the remaining 
   ``regular'' intervals}
  \label{Fig:chaos_regular}
\end{figure}

With Figure \ref{Fig:kappavar05}(c) we add a contour-plot for the
full 30-kyears simulation period, just to illustrate the sequence of 
17 breakdowns which were used for the numerical curve in 
Figure \ref{fig:bond_and_sim} (where the time direction 
is inverted, though). As seen in 
Figure \ref{fig:bond_and_sim}(b), those 17 events seem to obey
the same random walk law as the 54 Bond events, although 
our 30-kyears simulation time is still too short for 
a convincing statistics.
An interesting feature becomes visible in the wavelet spectrogram
of Figure \ref{Fig:kappavar05}(c) for $B(72^{\circ},t)$,
which shows the distribution of oscillations 
with period $\tau$ in the vicinity of time $t$.
During the breakdowns, characterized by a reduced energy 
in the 22.14-years 
period, we observe a significant increase of the energy 
in a wide range of $\tau$ from 30 to 1000-years. 
In fact the spectrum seems to become continuous here 
which 
reflects a transition to chaos due to nonlinearity. 
The quantitative difference between ``regular'' and ``chaotic''  
regimes is highlighted in Figure \ref{Fig:chaos_regular} 
which shows the wavelet spectral density 
integrated over the corresponding 
intervals, separated by the green dashed lines in Figure 
\ref{Fig:kappavar05}(d). 
With the energy in the $\tau$-range from 30 to 1000-years
being generally increased in the ``chaotic'' segments
by 1-2 orders of magnitude, the 
particular peak 
around 200-years is still markedly pronounced. 
A corresponding behaviour had been reported in Figure 6
of \cite{McCracken2013}.

In the Appendix, we will carry out a similar analysis as in 
Figure \ref{Fig:kappavar05} but for the case 
of fixed $\kappa=0.6$ without any time variation 
but stronger noise with $D=0.1$.  We will then also find breakdown
regions but no particular role of the Suess-de Vries cycle.

\section{Summary and open problems}

In this paper we have pursued the ambitious program 
of finding a  more or less complete and self-consistent 
description
of the most significant periodicities (and time-scales) 
of the solar dynamo. First, we recapitulated the basic idea
that the 22.14-years Hale cycle is synchronized by the
11.07-years tidal ($m=2$) forcing period of the tidally dominant 
VEJ-system,
the importance of which had been pointed out earlier by 
\cite{Hung2007,Scafetta2012,Wilson2013,Okhlopkov2014,Okhlopkov2016}.
In various numerical models 
\citep{Stefani2016,Stefani2018,Stefani2019} this 
(weak) tidal forcing was supposed to trigger 
a resonant excitation of 11.07-years oscillations of that 
part of the $\alpha$-effect which is related to 
a typical $m=1$ instability within or close to the 
tachocline,  such as the Tayler instability or a 
magneto-Rossby wave. 
Strong empirical evidence for a phase 
coherent 11.07-year Schwabe  cycle comes 
both from algae data in the early Holocene \citep{Vos2004}, 
and from $^{14}$C and $^{10}$B data for the 
last 600 years \citep{Stefani2020b}.
We note in passing that the numerical 
models produce, via the resonant dependence of the 
$\alpha$-effect on the magnetic field strength, also secondary 
peaks of solar activity, which might be linked to 
mid-term oscillations 
\citep{Obridko2007,Valdes2008,McIntosh2015,Bazilevskaya2016,Karak2018,Frick2020}.

Second, motivated by ideas of \cite{Cole1973,Wilson2013,Solheim2013}, we 
argued that  the Suess-de Vries cycle emerges as a 193-years beat 
period between the primary, 
tidally synchronized 22.14-years 
Hale cycle and the single strongest component of solar motion
around the barycenter of the planetary system, which is
governed by the 19.86-years synodic cycle of Jupiter and Saturn.
Without a detailed model for spin-orbit coupling at hand, we 
hypothesized that this coupling would lead to a periodic variation 
of the field loss parameter $\kappa$ in the tachocline region.
Numerically, the combination of such a $\kappa$-variation with the
synchronized component of the $\alpha$-effect led to  
a modulation of the dynamo wave with a beat period of 193 years, 
which manifests itself in a modulation of the North-South asymmetry
and, closely related to that, in a change of the dipole-quadrupole relation 
\citep{Knobloch1998,Moss2017} and the 
Gnevyshev-Ohl rule. We would like to point out
that the emergence of this beat period depends critically 
on the phase stability of the two underlying
11.07-years and 19.86-years processes, or, to put it otherwise: the 
existence of the long-term Suess-de Vries cycle gives
a ``backward argument'' for the synchronized character of the 
short-term Hale cycle.
As discussed in detail in \cite{Stefani2020a} (but only 
shortly touched upon in this paper), a stronger variation of 
$\kappa$ leads to the occurrence of a Wilson gap \citep{Wilson1987}
with two side peaks at 19.86 years and 25 years. Some aspects 
of this behaviour were illustrated in Figure \ref{Fig:period_vergleich}.

The main focus of this paper was, however, 
on the intermittent occurrence of grand minima, and clusters thereof.
We have observed irregular 
breakdowns of the 193-years modulated dynamo wave, 
preferably at instants where the North-South asymmetry 
reaches a certain critical level.
This threshold effect fits well to a corresponding observation of 
\cite{Tlatov2013} for the Maunder and Dalton minima.
Since those irregular breakdowns are already 
observed in the absence of any noise,
they seem to be connected with an intermittent transition to 
(deterministic) chaos, similar as in the supermodulation 
concept developed by \cite{Weiss2016}. 
For an appropriately chosen time-variation of $\kappa$ 
(and some weak noise), we obtained a waiting time distribution
with a similar mean value and standard deviation as 
inferred from the 54 Bond cycles 
observed over the last 80\,kyears. Such an intermittent 
transition to chaos would hamper any long-term predictability of 
solar activity (and the climatic changes connected with it), even 
if the planetary clocking of the shorter-term Hale and Suess-de Vries 
cycles could be confirmed. 

Based on a conventional $\alpha-\Omega$-dynamo, our
model thus required only the two synchronization periods 
11.07 years and 19.86 years, 
related to tidal forcing and the strongest component of 
solar orbital motion, respectively,
to  produce essentially all relevant periods, 
and ``periods'', of the solar dynamo. While this appears 
promising,  we conclude with a discussion of 
remaining problems and  ``missing links'' in the theory.

First, we have to admit that, up to present, the basic synchronization 
mechanism for the helicity of an $m=1$ mode by some tide-like ($m=2$) forcing 
has been evidenced only for the paradigmatic case
of a non-rotating, full cylinder \citep{Stefani2016}. 
Preliminary attempts for a hollow 
(more ``tachoclinic'') cylinder
were promising, although not entirely conclusive. Neither rotation nor 
stratification were implemented yet. A test of the same concept
in the simplified framework of an $m=1$ buoyancy instability of 
toroidal flux rings, as developed by \cite{Ferrizmas1994},
could  be very helpful and instructive in this respect.   
Interestingly, helicity synchronization with 
an $m=2$ forcing was numerically 
observed for the physically 
different, but topologically similar case of an $m=1$ large scale 
circulation of Rayleigh-B\'enard  convection in a cylinder 
\citep{Galindo2020}. 
A liquid metal 
experiment to confirm this effect is presently under 
way \citep{Stepanov2019,Juestel2020}. 

As this suggests a generic and robust character of the helicity 
synchronization mechanism, there is good hope to apply the same principle
also to the $m=1$ magneto-Rossby waves which were recently discussed  
by \cite{Dikpati2017,Marquez2017,McIntosh2017,Zaqarashvili2018}. Unstable shallow-water modes had 
been shown earlier \citep{Dikpati2003} 
to produce kinetic helicity, which is concentrated in the 
neighborhood of toroidal flux bands and migrates with them
toward the equator as the solar cycle progresses. 
It should not be too complicated to implement 
a tide-like $m=2$ forcing into those shallow-water models in
2D (here: in $\theta$ and longitude $\phi$) with the aim  
to identify a similar helicity synchronization mechanism as 
found for the Tayler instability.

As a next step one could think about combining such a 
2D model (in $\theta,\phi$) with another 2D model (in $\theta,r$) 
as it was utilized in various Babcock-Leighton type 
dynamo models \citep{Guerrero2007,Jouve2008}. Presently we are 
working on an enhancement of the latter type  of models 
in order to assess the direction of the butterfly diagram and to 
see whether a weak synchronized part
of $\alpha$ (with an amplitude of less then 1 m/s, as limited
by the argument of \cite{Opik1972}) is indeed sufficient for 
synchronizing the dynamo. 

Turning to the spin-orbit coupling based on the 19.86-years
orbital motion, and the resulting 193-years beat period, 
we first have to ask ourselves: does 
this beat period indeed correspond to the Suess-de Vries cycle? 
Since typical periods of this cycle 
between 190 until 210 years have been discussed in the 
literature, 193 years sounds not too bad in 
this respect. It also seems to be supported by a
recent result of \cite{Ma2020} who had found a 
195-years period for the strong and clearly expressed 
Suess-de Vries cycle in the relatively ``quiet''
interval between 800 and 1340 A.D.

But even if the equivalence of the  193-years modulation with the 
Suess-de Vries could once be confirmed, we would still be left with 
the problem to explain the coupling of the (mainly) 19.86-years periodic
orbit  into some internal, dynamo relevant motion. To the best of 
our knowledge, there have been 
no serious attempts to tackle this problem in its full beauty, 
including a realistic orbital motion, the 7 degree inclination of the 
Sun's rotation axis, and an alleged non-sphericity of the tachocline 
with a  
prolateness that might even vary with the magnetic field strength 
\citep{Dikpati2001}.
In this respect, we recall a relatively 
recent result on the similar problem of precession, for 
which a significant braking of the (solid body like)
rotational profile was observed already for weak precessional 
forcing \citep{Giesecke2018,Giesecke2019,Meunier2020}.
If such a braking effect would also occur in the solar 
spin-orbit coupling problem, it could indeed result in a 
change of the very sensitive adiabaticity \citep{Abreu2012}, and 
the associated $\kappa$ parameter as used here. 
Admittedly, this complex problem has to be left for future
studies.

With the main focus laid on the synchronized helicity of 
$m=1$ instabilities, we should not overlook alternative 
synchronization mechanisms based on axisymmetric ($m=0$) 
instabilities, the relevance of which 
had been discussed by several authors 
\citep{Dikpati2009,Rogers2011}. The recent detection of 
a double-diffusive helical magnetorotational instability 
for flows with positive shear  
\citep{Mama2019}
might be an interesting candidate in this respect, as it
depends quite sensitively on the ratio of toroidal to
poloidal field which, in turn, could be easily influenced
by variations of $\kappa$. 
 
In summary, given its skill in reproducing
the various solar cycles, and ``cycles'', with only mild 
(and indeed unessential) parameter fitting, 
our model looks like a reasonable choice for Occam's razor 
to point toward. Yet, we cannot completely 
rule out that we were maliciously mislead 
by all those regularities and coincidences, 
that in reality the Schwabe cycle results from a 
peculiar self-synchronization mechanism \citep{Hoyng1996}, and 
that less ``astrological'' explanations will also be 
found for the long-term rhythms of our home star.

\begin{acks}
This project has received funding 
from the European Research Council (ERC) under the 
European Union's Horizon 2020 research and innovation programme
(grant agreement No 787544).
The work was also supported in frame of the Helmholtz - RSF 
Joint Research Group ``Magnetohydrodynamic instabilities: Crucial 
relevance for large scale liquid metal batteries and
the sun-climate connection'', 
contract No HRSF-0044 and RSF-18-41-06201.
We thank Andr\'e Giesecke for a critical reading of
an early version of the manuscript. 
Inspiring discussions with  
J\"urg Beer, Antonio Ferriz Mas, Peter Frick, 
Rafael Rebolo,
G\"unther R\"udiger, Dmitry Sokoloff, Willie Soon, and 
Ian Wilson  on various aspects of the 
solar dynamo, and its synchronization, 
are gratefully acknowledged. 
\end{acks}

\section*{Disclosure of Potential Conflicts of Interest}
The authors declare that they have no conflicts of interest.

\section*{Appendix}

\begin{figure}[!ht]
\includegraphics[width=0.98\textwidth]{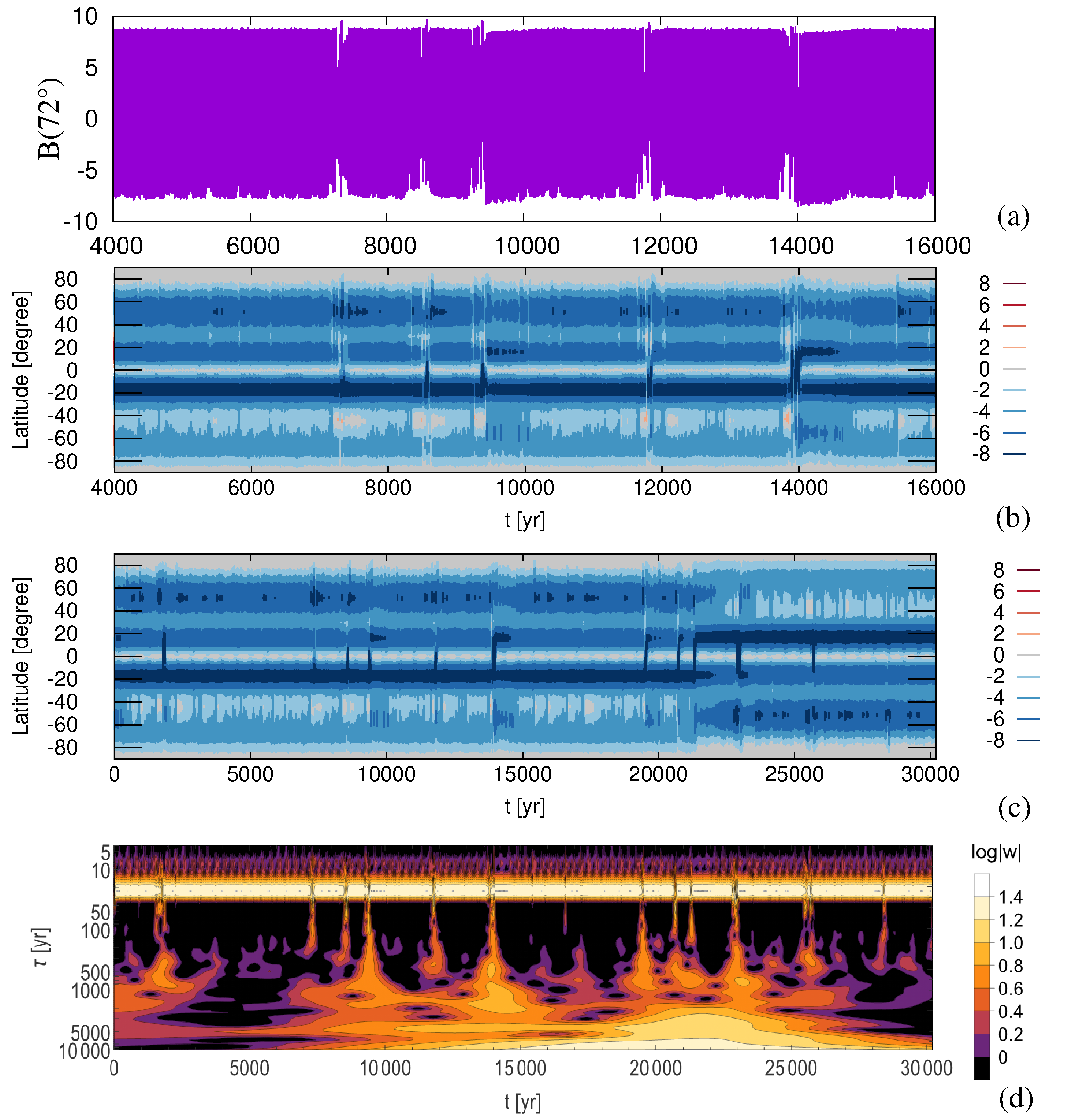}
\caption{(a) Behaviour of $B(72^{\circ},t)$ for $\omega_0=10000$, 
$\alpha^c_0=15$, $q^p_{\alpha}=0.2$, $q^c_{\alpha}=0.8$, $\alpha^p_0=50$, 
$D=0.1$ and $\kappa=0.6$, in the time interval
4000-16000 years.
(b) Contour-plot of $B(\theta,t)$ for the same parameters.
(c) Contour-plot for the full time interval
0-30200 years. (d) Wavelet diagram for the full time interval.}
\label{Fig:noisedetails}
\end{figure}

Figure \ref{Fig:noisedetails} shows, for all parameters 
kept as in Figure \ref{Fig:kappavar05}, except using 
 constant $\kappa=0.6$ and slightly increased $D=0.1$, 
 the occurrence
 of breakdowns under the exclusive influence of noise. 
 Compared to 
 Figure \ref{Fig:kappavar05} the behaviour seems to be different,
 with only one long-lasting change of the positive-negative and North-South
 asymmetry. 
 Not surprisingly for this case without $\kappa$ variation, there
 is no particular peak at 193 years in the wavelet spectrogram (d). 
 It seems that the transition to disorder/chaos is different in that
 case, and that the existence of a second frequency is decisive 
 in this respect.

\end{article} 


\begin{thebibliography}{59}
\ifx\bisbn     \undefined \def\bisbn  #1{ISBN #1}\fi
\ifx\binits    \undefined \def\binits#1{#1}\fi
\ifx\bauthor   \undefined \def\bauthor#1{#1}\fi
\ifx\batitle   \undefined \def\batitle#1{#1}\fi
\ifx\bjtitle   \undefined \def\bjtitle#1{\textit{#1}}\fi
\ifx\bvolume   \undefined \def\bvolume#1{\textbf{#1}}\fi
\ifx\byear     \undefined \def\byear#1{#1}\fi
\ifx\bissue    \undefined \def\bissue#1{#1}\fi
\ifx\bfpage    \undefined \def\bfpage#1{#1}\fi
\ifx\blpage    \undefined \def\blpage #1{#1}\fi
\ifx\burl      \undefined \def\burl#1{\textsf{#1}}\fi
\ifx\href      \undefined \def\href#1#2{\textsf{#2}}\fi
\ifx\betal     \undefined \def\betal{\textit{et al.}}\fi
\ifx\bctitle   \undefined \def\bctitle#1{#1}\fi
\ifx\beditor   \undefined \def\beditor#1{#1}\fi
\ifx\bbtitle   \undefined \def\bbtitle#1{\textit{#1}}\fi
\ifx\bedition  \undefined \def\bedition#1{#1}\fi
\ifx\bseriesno \undefined \def\bseriesno#1{\textbf{#1}}\fi
\ifx\blocation \undefined \def\blocation#1{#1}\fi
\ifx\bsertitle \undefined \def\bsertitle#1{\textit{#1}}\fi
\ifx\bsnm      \undefined \def\bsnm#1{#1}\fi
\ifx\bsuffix   \undefined \def\bsuffix#1{#1}\fi
\ifx\bparticle \undefined \def\bparticle#1{#1}\fi
\ifx\barticle  \undefined \def\barticle#1{}\fi
\ifx\binstitute  \undefined \def\binstitute#1{#1}\fi
\ifx\bpublisher  \undefined \def\bpublisher#1{#1}\fi
\ifx\doiurl    \undefined
  \def\doiurl#1{\href{http://dx.doi.org/#1}{\textsf{DOI}}}\fi
\ifx\arxivurl  \undefined
  \def\arxivurl#1{\href{http://arxiv.org/abs/#1}{\textsf{arXiv}}}\fi
\ifx\adsurl    \undefined
  \def\adsurl#1{\href{http://adsabs.harvard.edu/abs/#1}{\textsf{ADS}}}\fi
\ifx\botherref \undefined \def\botherref#1{}\fi
\ifx\url       \undefined \def\url#1{\textsf{#1}}\fi
\ifx\bchapter  \undefined \def\bchapter#1{}\fi
\ifx\bbook     \undefined \def\bbook#1{}\fi
\ifx\bcomment  \undefined \def\bcomment#1{#1}\fi
\ifx\oauthor   \undefined \def\oauthor#1{#1}\fi
\ifx\citeauthoryear \undefined\def \citeauthoryear#1{#1}\fi
\ifx\endbibitem\undefined \def\endbibitem{}\fi
\ifx\bconflocation  \undefined \def\bconflocation#1{#1} \fi



\bibitem[\protect\citeauthoryear{Abreu \textit{et~al.}}{{2012}}]{Abreu2012}
\begin{barticle}
\bauthor{\bsnm{Abreu}, \binits{J.A.}},
\bauthor{\bsnm{Beer}, \binits{J.}},
\bauthor{\bsnm{Ferriz-Mas}, \binits{A.}},
\bauthor{\bsnm{McCracken}, \binits{K.G.}},
\bauthor{\bsnm{Steinhilber}, \binits{F.}}:
\byear{{2012}},
\batitle{{Is there a planetary influence on solar activity?}}
\bjtitle{{Astron. Astrophys.}}
\bvolume{{548}},
\bfpage{{A88}}.
\doiurl{10.1051/0004-6361/201219997}.
\end{barticle}
\endbibitem



\bibitem[\protect\citeauthoryear{Arlt}{2009}]{Arlt2009}
\begin{barticle}
\bauthor{\bsnm{Arlt}, \binits{R.}}:
\byear{2009},
\batitle{The butterfly diagram in the eighteenth century}.
\bjtitle{Solar Phys.}
\bvolume{255},
\bfpage{143}.
\doiurl{10.1007/s11207-008-9306-5}.
\end{barticle}
\endbibitem




\bibitem[\protect\citeauthoryear{Bazilevskaya \textit{et~al.}}{2016}]{Bazilevskaya2016}
\begin{barticle}
\bauthor{\bsnm{Bazylevskaya}, \binits{G.A.}}
\bauthor{\bsnm{Kalinin}, \binits{M.S.}},
\bauthor{\bsnm{Krainev}, \binits{M.B.}}
\bauthor{\bsnm{Makhmutov}, \binits{V.S.}}
\bauthor{\bsnm{Svirzhevskaya}, \binits{A.K.}}
\bauthor{\bsnm{Svirzhevsky,}, \binits{N.S.}}
\bauthor{\bsnm{Stozhkov} \binits{Y.I.}}:
\byear{2016},
\batitle{On the relationship between quasi-biennial variations of solar activity, the
heliospheric magnetic field and cosmic rays}.
\bjtitle{Cosmic Res.}
\bvolume{54},
\bfpage{171}.
\doiurl{10.1134/S0010952516010019}.
\end{barticle}
\endbibitem


\bibitem[\protect\citeauthoryear{Beer, Tobias and Weiss}{1998}]{Beer1998}
\begin{barticle}
\bauthor{\bsnm{Beer}, \binits{J.}},
\bauthor{\bsnm{Tobias}, \binits{S.}},
\bauthor{\bsnm{Weiss}, \binits{N.}}:
\byear{1998},
\batitle{An active Sun throughout the Maunder minimum}.
\bjtitle{Solar Phys.}
\bvolume{181},
\bfpage{237}.
\doiurl{10.1023/A:1005026001784}.
\end{barticle}
\endbibitem



\bibitem[\protect\citeauthoryear{Bollinger}{1952}]{Bollinger1952}
\begin{barticle}
\bauthor{\bsnm{Bollinger}, \binits{C.J.}}:
\byear{1952},
\batitle{A 44.77 year Jupiter--Venus--Earth configuration Sun-tide period in
  solar-climatic cycles}.
\bjtitle{Proc. Okla. Acad. Sci.}
\bvolume{33},
\bfpage{307}.
\end{barticle}
\endbibitem

\bibitem[\protect\citeauthoryear{Bond \textit{et~al.}}{1997}]{Bond1997}
\begin{barticle}
\bauthor{\bsnm{Bond}, \binits{G.}},
\bauthor{\bsnm{Showers}, \binits{W.}},
\bauthor{\bsnm{Cheseby}, \binits{M.}},
\bauthor{\bsnm{Lotti}, \binits{R.}},
\bauthor{\bsnm{Almasi}, \binits{P.}},
\bauthor{\bsnm{deMenocal}, \binits{P.}},
\bauthor{\bsnm{Priore}, \binits{P.}},
\bauthor{\bsnm{Cullen}, \binits{H.}},
\bauthor{\bsnm{Haidas}, \binits{I.}},
\bauthor{\bsnm{Bonani}, \binits{G.}}:
\byear{1997},
\batitle{A pervasive millennial-scale cycle in North Atlantic Holocene and glacial climates}.
\bjtitle{Science.}
\bvolume{278},
\bfpage{1257}.
\doiurl{10.1126/science.278.5341.1257}.
\end{barticle}
\endbibitem

\bibitem[\protect\citeauthoryear{Bond \textit{et~al.}}{1999}]{Bond1999}
\begin{barticle}
\bauthor{\bsnm{Bond}, \binits{G.}},
\bauthor{\bsnm{Showers}, \binits{W.}},
\bauthor{\bsnm{Elliot}, \binits{M.}},
\bauthor{\bsnm{Evans}, \binits{M.}}
\bauthor{\bsnm{Lotti}, \binits{R.}},
\bauthor{\bsnm{Hajdas}, \binits{I.}},
\bauthor{\bsnm{Bonani}, \binits{G.}},
\bauthor{\bsnm{Johnson}, \binits{S.}}:
\byear{1999},
\batitle{The North Atlantic's 1-2 kyr climate rhythm: Relation to 
Heinrich events, Dansgard/Oeschger cycles and the Little Ice Age}.
\bjtitle{Mechanisms of Global Climate Change at Millennial Time Scales. Geophysical 
Monograph Series.}
\bvolume{112},
\bfpage{35}.
\doiurl{10.1029/GM112p0035}.
\end{barticle}
\endbibitem




\bibitem[\protect\citeauthoryear{Bond \textit{et~al.}}{2001}]{Bond2001}
\begin{barticle}
\bauthor{\bsnm{Bond}, \binits{G.}},
\bauthor{\bsnm{Kromer}, \binits{B.}},
\bauthor{\bsnm{Beer}, \binits{J.}},
\bauthor{\bsnm{Muscheler}, \binits{R.}},
\bauthor{\bsnm{Evans}, \binits{M.N.}},
\bauthor{\bsnm{Showers}, \binits{W.}},
\bauthor{\bsnm{Hoffmann}, \binits{S.}},
\bauthor{\bsnm{Lotti-Bond}, \binits{R.}},
\bauthor{\bsnm{Hajdas}, \binits{I.}},
\bauthor{\bsnm{Bonani}, \binits{G.}}:
\byear{2001},
\batitle{Persistent solar influence on North Atlantic climate during the Holocene}.
\bjtitle{Science.}
\bvolume{294},
\bfpage{2130}.
\doiurl{10.1126/science.1065680}.
\end{barticle}
\endbibitem

\bibitem[\protect\citeauthoryear{Braun \textit{et~al.}}{2005}]{Braun2004}
\begin{barticle}
\bauthor{\bsnm{Braun}, \binits{H.}},
\bauthor{\bsnm{Christl}, \binits{M.}},
\bauthor{\bsnm{Rahmstorf}, \binits{S.}},
\bauthor{\bsnm{Ganopolski}, \binits{A.}}
\bauthor{\bsnm{Mangini}, \binits{A.}},
\bauthor{\bsnm{Kubatzki}, \binits{C.}},
\bauthor{\bsnm{Roth}, \binits{K.}},
\bauthor{\bsnm{Kromer}, \binits{B.}}:
\byear{2004},
\batitle{Possible solar origin of the 1,470-year glacial climate cycle demonstrated in a coupled model}.
\bjtitle{Nature.}
\bvolume{438},
\bfpage{208}.
\doiurl{10.1038/nature04121}.
\end{barticle}
\endbibitem



\bibitem[\protect\citeauthoryear{Callebaut, de~Jager, and Duhau}{2012}]{Callebaut2012}
\begin{barticle}
\bauthor{\bsnm{Callebaut}, \binits{D.K.}},
\bauthor{\bparticle{de} \bsnm{Jager}, \binits{C.}},
\bauthor{\bsnm{Duhau}, \binits{S.}}:
\byear{2012},
\batitle{The influence of planetary attractions on the solar tachocline}.
\bjtitle{J. Atmos. Sol.-Terr. Phys.}
\bvolume{80},
\bfpage{73}.
\doiurl{10.1016/j.jastp.2012.03.005}.
\end{barticle}
\endbibitem




\bibitem[\protect\citeauthoryear{Charbonneau}{{2010}}]{Charbonneau2010}
\begin{botherref}
\oauthor{\bsnm{Charbonneau}, \binits{P.}}:
{2010},
{Dynamo models of the solar cycle}.
\textit{{Liv. Rev. Solar Phys.}}
\bvolume{7},
\bfpage{3}.
\doiurl{10.12942/lrsp-2010-3}.
\end{botherref}
\endbibitem


\bibitem[\protect\citeauthoryear{Charvatova}{1997}]{Charvatova1997}
\begin{barticle}
\bauthor{\bsnm{Charvatova}, \binits{I.}}:
\byear{1997},
\batitle{Solar-terrestrial and climatic phenomena in relation to solar inertial
  motion}.
\bjtitle{Surv. Geophys.}
\bvolume{18},
\bfpage{131}.
\doiurl{10.1023/A:1006527724221}.
\end{barticle}
\endbibitem


\bibitem[\protect\citeauthoryear{Cionco and Soon}{2015}]{Cionco2015}
\begin{barticle}
\bauthor{\bsnm{Cionco}, \binits{R.G.}},
\bauthor{\bsnm{Soon}, \binits{W.}}:
\byear{2015},
\batitle{A phenomenological study of the timing of solar
activity minima of the last millennium through a physical modeling
of the Sun-Planets interaction}.
\bjtitle{New Astron.}
\bvolume{34},
\bfpage{164}.
\doiurl{10.1016/j.newast.2014.07.001}.
\end{barticle}
\endbibitem


\bibitem[\protect\citeauthoryear{Cionco and Pavlov}{2018}]{Cionco2018}
\begin{barticle}
\bauthor{\bsnm{Cionco}, \binits{R.G.}},
\bauthor{\bsnm{Pavlov}, \binits{D.A.}}:
\byear{2018},
\batitle{Solar barycentric dynamics from a new solar-planetary ephemeris}.
\bjtitle{Astron. Astrophys.}
\bvolume{615},
\bfpage{A153}.
\doiurl{10.1051/0004-6361/201732349}.
\end{barticle}
\endbibitem



\bibitem[\protect\citeauthoryear{Cole}{1973}]{Cole1973}
\begin{barticle}
\bauthor{\bsnm{Cole}, \binits{T.W.}}:
\byear{1973},
\batitle{Periodicities in solar activity}.
\bjtitle{Solar Phys.}
\bvolume{30},
\bfpage{103}.
\doiurl{10.1007/BF00156178}.
\end{barticle}
\endbibitem

\bibitem[\protect\citeauthoryear{Condon and Schmidt}{1975}]{CondonSchmidt1975}
\begin{barticle}
\bauthor{\bsnm{Condon}, \binits{J.J.}},
\bauthor{\bsnm{Schmidt}, \binits{R.R.}}:
\byear{1975},
\batitle{Planetary tides and sunspot cycles}.
\bjtitle{Solar Phys.}
\bvolume{42},
\bfpage{529}.
\doiurl{10.1007/BF00149930}.
\end{barticle}
\endbibitem








\bibitem[\protect\citeauthoryear{De~Jager and Versteegh}{2005}]{DeJager2005}
\begin{barticle}
\bauthor{\bsnm{De~Jager}, \binits{C.}},
\bauthor{\bsnm{Versteegh}, \binits{G.}}:
\byear{2005},
\batitle{Do planetary motions drive solar variability?}
\bjtitle{Solar Phys.}
\bvolume{229},
\bfpage{175}.
\doiurl{10.1007/s11207-005-4086-7}.
\end{barticle}
\endbibitem

\bibitem[\protect\citeauthoryear{Dicke}{1978}]{Dicke1978}
\begin{barticle}
\bauthor{\bsnm{Dicke}, \binits{R.H.}}:
\byear{1978},
\batitle{Is there a chronometer hidden deep in the Sun?}
\bjtitle{Nature}
\bvolume{276},
\bfpage{676}.
\end{barticle}
\endbibitem

\bibitem[\protect\citeauthoryear{Dikpati and Gilman}{2001}]{Dikpati2001}
\begin{barticle}
\bauthor{\bsnm{Dikpati}, \binits{M.}},
\bauthor{\bsnm{Gilman}, \binits{P.A.}}:
\byear{2001},
\batitle{Prolateness of the solar tachocline inferred from latitudinal force
   balance in a magnetohydrodynamic shallow-water model}.
\bjtitle{Astrophys. J.}
\bvolume{552},
\bfpage{348}.
\doiurl{10.1086/320446}.
\end{barticle}
\endbibitem

\bibitem[\protect\citeauthoryear{Dikpati \textit{et~al.}}{2009}]{Dikpati2003}
\begin{barticle}
\bauthor{\bsnm{Dikpati}, \binits{M.}},
\bauthor{\bsnm{Gilman}, \binits{P.A}},
\bauthor{\bsnm{Rempel}, \binits{M.}}:
\byear{2003},
\batitle{Stability analysis of tachocline latitudinal differential
rotation and coexisting toroidal band using a shallow-water model}.
\bjtitle{Astrophys. J.}
\bvolume{596},
\bfpage{680}.
\doiurl{10.1086/377708}.
\end{barticle}
\endbibitem

\bibitem[\protect\citeauthoryear{Dikpati \textit{et~al.}}{2009}]{Dikpati2009}
\begin{barticle}
\bauthor{\bsnm{Dikpati}, \binits{M.}},
\bauthor{\bsnm{Gilman}, \binits{P.A}},
\bauthor{\bsnm{Cally}, \binits{P.S.}},
\bauthor{\bsnm{Miesch}, \binits{M.S.}}:
\byear{2009},
\batitle{Axisymmetric MHD instabilities in solar/stellar tachoclines}.
\bjtitle{Astrophys. J.}
\bvolume{692},
\bfpage{1421}.
\doiurl{10.1088/0004-637X/692/2/1421}.
\end{barticle}
\endbibitem



\bibitem[\protect\citeauthoryear{Dikpati \textit{et~al.}}{2017}]{Dikpati2017}
\begin{barticle}
\bauthor{\bsnm{Dikpati}, \binits{M.}},
\bauthor{\bsnm{Cally}, \binits{P.S.}},
\bauthor{\bsnm{McIntosh}, \binits{S.W.}},
\bauthor{\bsnm{Heifetz}, \binits{E.}}:
\byear{2017},
\batitle{The origin of the ``Seasons'' in Space Weather}.
\bjtitle{Sci. Rep.}
\bvolume{7},
\bfpage{14750}.
\doiurl{10.1038/s41598-017-14957-x}.
\end{barticle}
\endbibitem

\bibitem[\protect\citeauthoryear{Dima and Lohmann}{2009}]{Dima2009}
\begin{barticle}
\bauthor{\bsnm{Dima}, \binits{M.}},
\bauthor{\bsnm{Lohmann}, \binits{G.}}:
\byear{2009},
\batitle{Conceptual model for millennial climate
variability: a possible combined solar-thermohaline circulation
origin for the 1,500-year cycle}.
\bjtitle{Clim. Dyn.}
\bvolume{32},
\bfpage{301}.
\doiurl{10.1007/s00382-008-0471-x}.
\end{barticle}
\endbibitem


\bibitem[\protect\citeauthoryear{Fairbridge and Shirley}{1987}]{Fairbridge1987}
\begin{barticle}
\bauthor{\bsnm{Fairbridge}, \binits{R.W.}},
\bauthor{\bsnm{Shirley}, \binits{J.H.}}:
\byear{1987},
\batitle{Prolonged minima and the 179-yr cycle of the solar 
inertial motions}.
\bjtitle{Solar Phys.}
\bvolume{110},
\bfpage{191}.
\doiurl{10.1007/BF00148211}.
\end{barticle}
\endbibitem




\bibitem[\protect\citeauthoryear{Ferriz~Mas, Schmitt, and
  Sch\"ussler}{1994}]{Ferrizmas1994}
\begin{barticle}
\bauthor{\bsnm{Ferriz~Mas}, \binits{A.}},
\bauthor{\bsnm{Schmitt}, \binits{D.}},
\bauthor{\bsnm{Sch\"ussler}, \binits{M.}}:
\byear{1994},
\batitle{A dynamo effect due to instability of magnetic flux tubes}.
\bjtitle{Astron. Astrophys.}
\bvolume{289},
\bfpage{949}.
\end{barticle}
\endbibitem

\bibitem[\protect\citeauthoryear{Folkner \textit{et~al.}}{2014}]{Folkner2014}
\begin{barticle}
\bauthor{\bsnm{Folkner}, \binits{W.M.}},
\bauthor{\bsnm{Williams}, \binits{J.G.}},
\bauthor{\bsnm{Boggs}, \binits{D.H.}},
\bauthor{\bsnm{Park}, \binits{R.S.}},
\bauthor{\bsnm{Kuchynka}, \binits{P.}}:
\byear{2014},
\batitle{The planetary and lunar ephemerides DE430 and DE431}.
\bjtitle{IPN Progress Report}
\bvolume{42-196},
\bfpage{1}.
\end{barticle}
\endbibitem


\bibitem[\protect\citeauthoryear{Frick \textit{et~al.}}{2020}]{Frick2020}
\begin{barticle}
\bauthor{\bsnm{Frick}, \binits{P.}},
\bauthor{\bsnm{Sokoloff}, \binits{D.}},
\bauthor{\bsnm{Stepanov}, \binits{R.}},
\bauthor{\bsnm{Pipin}, \binits{V.}},
\bauthor{\bsnm{Usoskin}, \binits{I.}}:
\byear{2020},
\batitle{Spectral characteristic of mid-term quasi-periodicities in sunspot data}.
\bjtitle{Mon. Not. R. Astron. Soc.}
\bvolume{491},
\bfpage{5572}.
\doiurl{10.1093/mnras/stz3238}.
\end{barticle}
\endbibitem

\bibitem[\protect\citeauthoryear{Galindo}{2020}]{Galindo2020}
\begin{barticle}
\bauthor{\bsnm{Galindo}, \binits{V.}}:
\byear{2020},
\batitle{personal communication}.
\end{barticle}
\endbibitem


\bibitem[\protect\citeauthoryear{Guerrero and de Gouveia Dal Pino}{2007}]{Guerrero2007}
\begin{barticle}
\bauthor{\bsnm{Guerrero}, \binits{G.}},
\bauthor{\bsnm{de Gouveia Dal Pino}, \binits{E.M.}}:
\byear{2007},
\batitle{How does the shape and thickness of the tachocline affect the distribution of the toroidal magnetic fields in the solar dynamo?}.
\bjtitle{Astron. Astrophys.}
\bvolume{464},
\bfpage{341}.
\doiurl{10.1051/0004-6361:20065834}.
\end{barticle}
\endbibitem


\bibitem[\protect\citeauthoryear{Giesecke \textit{et~al.}}{2018}]{Giesecke2018}
\begin{barticle}
\bauthor{\bsnm{Giesecke}, \binits{A.}},
\bauthor{\bsnm{Vogt}, \binits{T.}},
\bauthor{\bsnm{Gundrum}, \binits{T.}},
\bauthor{\bsnm{Stefani}, \binits{F.}}:
\byear{2018},
\batitle{Nonlinear large scale flow in a precessing cylinder and its ability to drive 
dynamo action}.
\bjtitle{Phys. Rev. Lett.}
\bvolume{120},
\bfpage{024502}.
\doiurl{10.1103/PhysRevLett.120.024502}.
\end{barticle}
\endbibitem


\bibitem[\protect\citeauthoryear{Giesecke \textit{et~al.}}{2019}]{Giesecke2019}
\begin{barticle}
\bauthor{\bsnm{Giesecke}, \binits{A.}},
\bauthor{\bsnm{Vogt}, \binits{T.}},
\bauthor{\bsnm{Gundrum}, \binits{T.}}:
\bauthor{\bsnm{Stefani}, \binits{F.}}:
\byear{2019},
\batitle{Kinematic dynamo action of a precession-driven flow based on the results of water experiments and hydrodynamic simulations.}.
\bjtitle{Geophys. Astrophys. Fluid Dyn.}
\bvolume{113},
\bfpage{235}.
\doiurl{10.1080/03091929.2018.1506774}.
\end{barticle}
\endbibitem



\bibitem[\protect\citeauthoryear{Gnevyshev and Ohl}{1948}]{Gnevyshev1948}
\begin{barticle}
\bauthor{\bsnm{Gnevyshev}, \binits{M.N.}},
\bauthor{\bsnm{Ohl}, \binits{A.I.}}:
\byear{1948},
\batitle{On the 22-year cycle of solar activity}.
\bjtitle{Astron. J.}
\bvolume{25},
\bfpage{18-20}.
\end{barticle}
\endbibitem


\bibitem[\protect\citeauthoryear{Hathaway}{2010}]{Hathaway2010}
\begin{barticle}
\bauthor{\bsnm{Hathaway}, \binits{D.H.}}:
\byear{2015},
\batitle{The solar cycle}.
\bjtitle{Liv. Rev. Sol. Phys.}
\bvolume{12},
\bfpage{4}.
\doiurl{10.12942/lrsp-2015-4}.
\end{barticle}
\endbibitem



\bibitem[\protect\citeauthoryear{Howe}{2009}]{Howe2009}
\begin{barticle}
\bauthor{\bsnm{Howe}, \binits{R.}}:
\byear{2009},
\batitle{Solar interior rotation and its variation}.
\bjtitle{Liv. Rev. Sol. Phys.}
\bvolume{6},
\bfpage{1}.
\doiurl{10.12942/lrsp-2009-1}.
\end{barticle}
\endbibitem

\bibitem[\protect\citeauthoryear{Hoyng}{1996}]{Hoyng1996}
\begin{barticle}
\bauthor{\bsnm{Hoyng}, \binits{P.}}:
\byear{1996},
\batitle{Is the solar cycle timed by a clock?}
\bjtitle{Solar Phys.}
\bvolume{169},
\bfpage{253}.
\doiurl{10.1007/BF00190603}.
\end{barticle}
\endbibitem

\bibitem[\protect\citeauthoryear{Hung}{2007}]{Hung2007}
\begin{barticle}
\bauthor{\bsnm{Hung}, \binits{C.-C.}}:
\byear{2007},
\batitle{Apparent relations between solar activity and solar tides caused by
  the planets.}
\bfpage{NASA/TM-2007-214817}.
\end{barticle}
\endbibitem

\bibitem[\protect\citeauthoryear{Jennings and Weiss}{1991}]{Jennings1991}
\begin{barticle}
\bauthor{\bsnm{Jennings}, \binits{R.L.}},
\bauthor{\bsnm{Weiss}, \binits{N.O.}}:
\byear{1991},
\batitle{Symmetry breaking in stellar dynamos}.
\bjtitle{Mon. Not. R. Astr. Soc.}
\bvolume{252},
\bfpage{249}.
\doiurl{10.1093/mnras/252.2.249}.
\end{barticle}
\endbibitem

\bibitem[\protect\citeauthoryear{Jones}{1983}]{Jones1983}
\begin{barticle}
\bauthor{\bsnm{Jones}, \binits{C.A.}}:
\byear{1983},
\batitle{Model equations for the solar dynamo.
In: Soward, A.M. (ed.) {\it Stellar and Planetary Magnetism}, 
Gordon and Breach, New York},
\bfpage{193}.
\end{barticle}
\endbibitem


\bibitem[\protect\citeauthoryear{Jose}{1965}]{Jose1965}
\begin{barticle}
\bauthor{\bsnm{Jose}, \binits{P.D.}}:
\byear{1965},
\batitle{Sun's motion and sunspots}.
\bjtitle{Astron. J.}
\bvolume{70},
\bfpage{193}.
\doiurl{10.1086/109714}.
\end{barticle}
\endbibitem

\bibitem[\protect\citeauthoryear{Jouve \textit{et~al.}}{2008}]{Jouve2008}
\begin{barticle}
\bauthor{\bsnm{Jouve}, \binits{L.}},
\bauthor{\bsnm{Brun}, \binits{A.S.}},
\bauthor{\bsnm{Arlt}, \binits{R.}},
\bauthor{\bsnm{Brandenburg}, \binits{A.}},
\bauthor{\bsnm{Dikpati}, \binits{M.}},
\bauthor{\bsnm{Bonanno}, \binits{A.}},
\bauthor{\bsnm{K\"apyl\"a}, \binits{P.J.}},
\bauthor{\bsnm{Moss}, \binits{D.}},
\bauthor{\bsnm{Rempel}, \binits{M.}},
\bauthor{\bsnm{Gilman}, \binits{P.}},
\bauthor{\bsnm{Korpi}, \binits{M.J.}},
\bauthor{\bsnm{Kosovichev}, \binits{A.G.}}:
\byear{2015},
\batitle{A solar mean field dynamo benchmark}.
\bjtitle{Astron. Astrophys.}
\bvolume{483},
\bfpage{949}.
\doiurl{10.1051/0004-6361:20078351}.
\end{barticle}
\endbibitem


\bibitem[\protect\citeauthoryear{J\"ustel \textit{et~al.}}{2020}]{Juestel2020}
\begin{barticle}
\bauthor{\bsnm{J\"ustel}, \binits{P.}},
\bauthor{\bsnm{R\"ohrborn}, \binits{S.}},
\bauthor{\bsnm{Frick}, \binits{P.}},
\bauthor{\bsnm{Galindo}, \binits{V.}},
\bauthor{\bsnm{Gundrum}, \binits{T.}},
\bauthor{\bsnm{Schindler}, \binits{F.}},
\bauthor{\bsnm{Stefani}, \binits{F.}},
\bauthor{\bsnm{Stepanov}, \binits{R.}},
\bauthor{\bsnm{Vogt}, \binits{T.}}:
\byear{2020},
\batitle{Generating a tide-like flow in a cylindrical vessel by electromagnetic
forcing}.
\bjtitle{Phys. Fluids (submitted)},
\bfpage{arXiv:2006.03491}.
\end{barticle}
\endbibitem





\bibitem[\protect\citeauthoryear{Karak, Mandal and Banarjee}{2018}]{Karak2018}
\begin{barticle}
\bauthor{\bsnm{Karak}, \binits{B.B.}},
\bauthor{\bsnm{Mandal}, \binits{S.}},
\bauthor{\bsnm{Banarjee}, \binits{D.}}:
\byear{2018},
\batitle{Double-peaks of the solar cycle: an explanation from a dynamo
model}.
\bjtitle{Astrophys. J.}
\bvolume{866},
\bfpage{17}.
\doiurl{10.3847/1538-4357/aada0d}.
\end{barticle}
\endbibitem


\bibitem[\protect\citeauthoryear{Knobloch, Tobias and Weiss}{1998}]{Knobloch1998}
\begin{barticle}
\bauthor{\bsnm{Knobloch}, \binits{E}},
\bauthor{\bsnm{Tobias}, \binits{S.M.}},
\bauthor{\bsnm{Weiss}, \binits{N.O.}}:
\byear{1998},
\batitle{Modulation and symmetry changes in stellar dynamos}.
\bjtitle{Mon. Not. R. Astron. Soc.}
\bvolume{297},
\bfpage{1123}.
\doiurl{10.1046/j.1365-8711.1998.01572.x}.
\end{barticle}
\endbibitem



\bibitem[\protect\citeauthoryear{Kotov and Haneychuk}{2020}]{Kotov2020}
\begin{barticle}
\bauthor{\bsnm{Kotov}, \binits{V.A.}},
\bauthor{\bsnm{Haneychuk}, \binits{V.I.}}:
\byear{2020},
\batitle{Oscillations of solar photosphere: 45 years of observations}.
\bjtitle{Astro. Nachr.}
\doiurl{10.1002/asna.202013797}.
\end{barticle}
\endbibitem


\bibitem[\protect\citeauthoryear{Kudryavtsev and Dergachev}{2020}]{Kudryavtsev2020}
\begin{barticle}
\bauthor{\bsnm{Kudryavtsev}, \binits{I.V.}},
\bauthor{\bsnm{Dergachev}, \binits{V.A.}}:
\byear{2020},
\batitle{Reconstruction of heliospheric modulation potential based on 
radiocarbon data in the time interval 17000-5000 years B.C.}.
\bjtitle{Geomagn. Aeron.}
\bvolume{59},
\bfpage{1099}.
\doiurl{10.1134/S0016793219080115}.
\end{barticle}
\endbibitem



\bibitem[\protect\citeauthoryear{Kuzanyan and Sokoloff}{1997}]{Kuzanyan1997}
\begin{barticle}
\bauthor{\bsnm{Kuzanyan}, \binits{K.M.}},
\bauthor{\bsnm{Sokoloff}, \binits{D.}}:
\byear{1997},
\batitle{Half-width of a solar dynamo wave in Parker's migratory dynamo}.
\bjtitle{Solar Phys.}
\bvolume{173},
\bfpage{1}.
\doiurl{10.1023/A:1004983000503 }.
\end{barticle}
\endbibitem



\bibitem[\protect\citeauthoryear{Landscheidt}{1999}]{Landscheidt1999}
\begin{barticle}
\bauthor{\bsnm{Landscheidt}, \binits{T.}}:
\byear{1999},
\batitle{Extrema in sunspot cycle linked to Sun's motion}.
\bjtitle{Solar Phys.}
\bvolume{189},
\bfpage{413}.
\doiurl{10.1023/A:1005287705442}.
\end{barticle}
\endbibitem

\bibitem[\protect\citeauthoryear{L\"udecke, Weiss and Hempelmann}{2015}]{Luedecke2015}
\begin{barticle}
\bauthor{\bsnm{L\"udecke}, \binits{H.-J.}},
\bauthor{\bsnm{Weiss}, \binits{C.O.}},
\bauthor{\bsnm{Hempelmann}, \binits{A.}}:
\byear{2015},
\batitle{Paleoclimate forcing by the solar De Vries/Suess cycle}.
\bjtitle{Clim. Past Discuss.}
\bvolume{11},
\bfpage{279}.
\doiurl{10.5194/cpd-11-279-2015}.
\end{barticle}
\endbibitem


\bibitem[\protect\citeauthoryear{Ma and Vaquero}{2020}]{Ma2020}
\begin{barticle}
\bauthor{\bsnm{Ma}, \binits{L.}},
\bauthor{\bsnm{Vaquero}, \binits{J.M.}},
\byear{2020}
\batitle{New evidence of the Suess/de Vries cycle existing in historical naked-eye 
observations of sunspots}.
\bjtitle{Open Astron.}
\bvolume{29},
\bfpage{28}.
\doiurl{10.1515/astro-2020-0004}.
\end{barticle}
\endbibitem



\bibitem[\protect\citeauthoryear{Mamatsashvili \textit{et~al.}}{2019}]{Mama2019}
\begin{barticle}
\bauthor{\bsnm{Mamatsashvili}, \binits{G.}},
\bauthor{\bsnm{Stefani}, \binits{F.}},
\bauthor{\bsnm{Hollerbach}, \binits{R.}},
\bauthor{\bsnm{R\"udiger}, \binits{G.}}:
\byear{2019},
\batitle{Two types of axisymmetric helical magnetorotational 
instability in rotating flows with positive shear}.
\bjtitle{Phys. Rev. Fluids},
\bvolume{4},
\bfpage{103905}.
\doiurl{10.1103/PhysRevFluids.4.103905}.
\end{barticle}
\endbibitem



\bibitem[\protect\citeauthoryear{Marquez-Artavia, Jones, and Tobias,}{2017}]{Marquez2017}
\begin{barticle}
\bauthor{\bsnm{Marquez-Artavia}, \binits{X.}},
\bauthor{\bsnm{Jones}, \binits{C.A.}},
\bauthor{\bsnm{Tobias}, \binits{S.M.}}:
\byear{2017},
\batitle{Rotating magnetic shallow water waves and instabilities in a sphere}.
\bjtitle{Geophys. Astrophys. Fluid Dyn.}
\bvolume{111},
\bfpage{282}.
\doiurl{10.1080/03091929.2017.1301937}.
\end{barticle}
\endbibitem


\bibitem[\protect\citeauthoryear{Mayewski \textit{et~al.}}{1997}]{Mayewski1997}
\begin{barticle}
\bauthor{\bsnm{Mayewski}, \binits{P.A.}},
\bauthor{\bsnm{Meeker}, \binits{L.D.}},
\bauthor{\bsnm{Twickler}, \binits{M.S.}},
\bauthor{\bsnm{Whitlow}, \binits{S.}},
\bauthor{\bsnm{Yang}, \binits{Q.}},
\bauthor{\bsnm{Berry Lyons}, \binits{W.}},
\bauthor{\bsnm{Prentice}, \binits{M.}}.
\byear{1997},
\batitle{Major features and forcing of high-latitude northern 
hemisphere atmospheric circulation using a 111,000-year-long glaciochemical series}.
\bjtitle{J. Geophys. Res.: Oceans}
\bvolume{102},
\bfpage{26345}.
\doiurl{10.1029/96JC03365}.
\end{barticle}
\endbibitem


\bibitem[\protect\citeauthoryear{McCracken, Beer and Steinhilber}{2013}]{McCracken2013}
\begin{barticle}
\bauthor{\bsnm{McCracken}, \binits{K.G.}},
\bauthor{\bsnm{Beer}, \binits{J.}},
\bauthor{\bsnm{Steinhilber}, \binits{F.}},
\bauthor{\bsnm{Abreu}, \binits{J.}}:
\byear{2013},
\batitle{A phenomenological study of the cosmic ray variations 
over the Past 9400 years, and their implications regarding solar activity and the solar dynamo}.
\bjtitle{Solar Phys.}
\bvolume{286},
\bfpage{609}.
\doiurl{10.1007/s11207-013-0265-0}.
\end{barticle}
\endbibitem

\bibitem[\protect\citeauthoryear{McCracken, Beer and Steinhilber}{2014}]{McCracken2014}
\begin{barticle}
\bauthor{\bsnm{McCracken}, \binits{K.G.}},
\bauthor{\bsnm{Beer}, \binits{J.}},
\bauthor{\bsnm{Steinhilber}, \binits{F.}}:
\byear{2014},
\batitle{Evidence for planetary forcing of the cosmic ray intensity and solar
activity throughout the past 9400 years}.
\bjtitle{Solar Phys.}
\bvolume{289},
\bfpage{3207}.
\doiurl{10.1007/s11207-014-0510-1}.
\end{barticle}
\endbibitem



\bibitem[\protect\citeauthoryear{McIntosh \textit{et~al.}}{2015}]{McIntosh2015}
\begin{barticle}
\bauthor{\bsnm{McIntosh}, \binits{S.W. et al.}}:
\byear{2015},
\batitle{The solar magnetic activity band 
interaction and instabilities that shape quasi-periodic variability}.
\bjtitle{Nature Comm.}
\bvolume{6},
\bfpage{6491}.
\doiurl{10.1038/ncomms7491}.
\end{barticle}
\endbibitem

\bibitem[\protect\citeauthoryear{McIntosh \textit{et~al.}}{2017}]{McIntosh2017}
\begin{barticle}
\bauthor{\bsnm{McIntosh}, \binits{S.W.}},
\bauthor{\bsnm{Cramer}, \binits{W.J.}},
\bauthor{\bsnm{Pichardo Marcano}, \binits{M}},
\bauthor{\bsnm{Leamon}, \binits{R.J.}}:
\byear{2017},
\batitle{The detection of Rossby-like waves on the Sun}.
\bjtitle{Nature Astron.}
\bvolume{1},
\bfpage{0086}.
\doiurl{10.1038/s41550-017-0086}.
\end{barticle}
\endbibitem

\bibitem[\protect\citeauthoryear{Meunier and Albrecht}{2020}]{Meunier2020}
\begin{barticle}
\bauthor{\bsnm{Meunier}, \binits{P.}},
\bauthor{\bsnm{Albrecht}, \binits{T.}}:
\byear{2020},
\bfpage{personal communication}.
\end{barticle}
\endbibitem


\bibitem[\protect\citeauthoryear{Moss and Sokoloff}{2017}]{Moss2017}
\begin{barticle}
\bauthor{\bsnm{Moss}, \binits{D.L.}},
\bauthor{\bsnm{Sokoloff}, \binits{D.}}:
\byear{2017},
\batitle{Parity fluctuations in stellar dynamos}.
\bjtitle{Astron. Rep.}
\bvolume{61},
\bfpage{878}.
\doiurl{10.1134/S1063772917100079}.
\end{barticle}
\endbibitem


\bibitem[\protect\citeauthoryear{Muscheler \textit{et~al.}}{2007}]{Muscheler2007}
\begin{barticle}
\bauthor{\bsnm{Muscheler}, \binits{R.}}
\bauthor{\bsnm{Joos}, \binits{F.}}
\bauthor{\bsnm{Beer}, \binits{J.}}
\bauthor{\bsnm{M\"uller}, \binits{S.A.}}
\bauthor{\bsnm{Vonmoos}, \binits{M.}},
\bauthor{\bsnm{Snowball}, \binits{I.}}:
\byear{2007},
\batitle{Solar activity during the last 1000 yr inferred from 
radionuclide records}.
\bjtitle{Quat. Sci. Rev.,}
\bvolume{26},
\bfpage{82}.
\doiurl{10.1016/j.quascirev.2006.07.012}.
\end{barticle}
\endbibitem




\bibitem[\protect\citeauthoryear{Obridko and Shelting}{2007}]{Obridko2007}
\begin{barticle}
\bauthor{\bsnm{Obridko}, \binits{V.N.}},
\bauthor{\bsnm{Shelting}, \binits{B.D.}}:
\byear{2007},
\batitle{Occurrence of the 1.3-year periodicity in the 
large-scale solar magnetic field for 8 solar cycles}.
\bjtitle{Adv. Space Res.}
\bvolume{40},
\bfpage{1006}.
\doiurl{10.1016/j.asr.2007.04.105}.
\end{barticle}
\endbibitem


\bibitem[\protect\citeauthoryear{Okhlopkov}{2014}]{Okhlopkov2014}
\begin{barticle}
\bauthor{\bsnm{Okhlopkov}, \binits{V.P.}}:
\byear{2014},
\batitle{The 11-year cycle of solar activity and configurations of the
  planets}.
\bjtitle{Mosc. Univ. Phys. B.}
\bvolume{69},
\bfpage{257}.
\doiurl{10.3103/S0027134914030126}.
\end{barticle}
\endbibitem


\bibitem[\protect\citeauthoryear{Okhlopkov}{2016}]{Okhlopkov2016}
\begin{barticle}
\bauthor{\bsnm{Okhlopkov}, \binits{V.P.}}:
\byear{2016},
\batitle{The gravitational influence of Venus, the Earth, and Jupiter on the
   11-year cycle of solar activity}.
\bjtitle{Mosc. Univ. Phys. B.}
\bvolume{71},
\bfpage{440}.
\doiurl{10.3103/S0027134916040159}.
\end{barticle}
\endbibitem






\bibitem[\protect\citeauthoryear{\"Opik}{1972}]{Opik1972}
\begin{barticle}
\bauthor{\bsnm{\"Opik}, \binits{E.}}:
\byear{1972},
\batitle{Solar-planetary tides and sunspots}.
\bjtitle{I. Astron. J.}
\bvolume{10},
\bfpage{298}.
\end{barticle}
\endbibitem


\bibitem[\protect\citeauthoryear{Palus \textit{et~al.}}{2000}]{Palus2000}
\begin{barticle}
\bauthor{\bsnm{Palus}, \binits{M.}},
\bauthor{\bsnm{Kurths}, \binits{J.}},
\bauthor{\bsnm{Schwarz}, \binits{U.}},
\bauthor{\bsnm{Novotna}, \binits{D.}},
\bauthor{\bsnm{Charvatova}, \binits{I.}}:
\byear{2000},
\batitle{Is the solar activity cycle synchronized with the solar inertial
  motion?}
\bjtitle{Int. J. Bifurc. Chaos}
\bvolume{10},
\bfpage{2519}.
\doiurl{10.1142/S0218127400001766}.
\end{barticle}
\endbibitem

\bibitem[\protect\citeauthoryear{Parker}{1955}]{Parker1955}
\begin{barticle}
\bauthor{\bsnm{Parker}, \binits{E.N.}}:
\byear{1955},
\batitle{Hydromagnetic dynamo models}.
\bjtitle{Astrophys. J.}
\bvolume{122},
\bfpage{293}.
\doiurl{10.1086/146087}.
\end{barticle}
\endbibitem



\bibitem[\protect\citeauthoryear{Pitts and Tayler}{1985}]{Pittstayler1985}
\begin{barticle}
\bauthor{\bsnm{Pitts}, \binits{E.}},
\bauthor{\bsnm{Tayler}, \binits{R.J.}}:
\byear{1985},
\batitle{The adiabatic stability of stars containing magnetic-fields. 6. The
  influence of rotation}.
\bjtitle{Mon. Not. Roy. Astron. Soc.}
\bvolume{216},
\bfpage{139}.
\doiurl{10.1093/mnras/216.2.139}.
\end{barticle}
\endbibitem


\bibitem[\protect\citeauthoryear{Proctor}{2007}]{Proctor2007}
\begin{barticle}
\bauthor{\bsnm{Proctor}, \binits{M.R.E.}}:
\byear{2007},
\batitle{Effects of fluctuations on $\alpha \Omega$ dynamo models}.
\bjtitle{Mon. Not. R. Astron. Soc.}
\bvolume{382},
\bfpage{L39}.
\doiurl{10.1111/j.1745-3933.2007.00385.x}.
\end{barticle}
\endbibitem


\bibitem[\protect\citeauthoryear{Raynaud and Tobias}{2016}]{Raynaud2016}
\begin{barticle}
\bauthor{\bsnm{Raynaud}, \binits{R.}},
\bauthor{\bsnm{Tobias}, \binits{S.M.}}:
\byear{2016},
\batitle{Convective dynamo action in a spherical shell: symmetries and modulation}.
\bjtitle{J. Fluid Mech.}
\bvolume{799},
\bfpage{R6}.
\doiurl{10.1017/jfm.2016.407}.
\end{barticle}
\endbibitem



\bibitem[\protect\citeauthoryear{Roald and Thomas}{1997}]{Roald1997}
\begin{barticle}
\bauthor{\bsnm{Roald}, \binits{C.B.}},
\bauthor{\bsnm{Thomas}, \binits{J.H.}}:
\byear{1997},
\batitle{Simple solar dynamo models with variable $\alpha$ and $\omega$ effects}.
\bjtitle{Mon. Not. R. Astron. Soc.}
\bvolume{288},
\bfpage{551}.
\doiurl{10.1093/mnras/288.3.551}.
\end{barticle}
\endbibitem


\bibitem[\protect\citeauthoryear{Rogers}{2011}]{Rogers2011}
\begin{barticle}
\bauthor{\bsnm{Rogers}, \binits{T.M.}}:
\byear{2011},
\batitle{Toroidal field reversals and the axisymmetric Tayler instability}.
\bjtitle{Mon. Not. R. Astron. Soc.}
\bvolume{288},
\bfpage{551}.
\doiurl{10.1088/0004-637X/735/2/100}.
\end{barticle}
\endbibitem


\bibitem[\protect\citeauthoryear{R\"udiger and Schultz}{2020}]{Ruediger2020}
\begin{barticle}
\bauthor{\bsnm{R\"udiger}, \binits{G.}},
\bauthor{\bsnm{Schultz}, \binits{M.}}:
\byear{2020},
\batitle{Large-scale dynamo action of magnetized Taylor-Couette flows}.
\bjtitle{Mon. Not. R. Astron. Soc.}
\bvolume{493},
\bfpage{1249}.
\doiurl{10.1093/mnras/staa293}.
\end{barticle}
\endbibitem



\bibitem[\protect\citeauthoryear{Scafetta}{2012}]{Scafetta2012}
\begin{barticle}
\bauthor{\bsnm{Scafetta}, \binits{N.}}:
\byear{2012},
\batitle{Does the Sun work as a nuclear fusion amplifier of planetary tidal 
forcing? A proposal for a physical mechanism based on the mass-luminosity relation}.
\bjtitle{J. Atmos. Sol.-Terr. Phys.}
\bvolume{81-82},
\bfpage{27}.
\doiurl{10.1016/j.jastp.2012.04.002}.
\end{barticle}
\endbibitem



\bibitem[\protect\citeauthoryear{Scafetta \textit{et~al.}}{2016}]{Scafetta2016}
\begin{barticle}
\bauthor{\bsnm{Scafetta}, \binits{N.}},
\bauthor{\bsnm{Milani}, \binits{F.}},
\bauthor{\bsnm{Bianchini}, \binits{A.}},
\bauthor{\bsnm{Ortolani}, \binits{S.}}:
\byear{2016},
\batitle{On the astronomical origin of the Hallstatt oscillation found in
   radiocarbon and climate records throughout the Holocene}.
\bjtitle{Earth Sci. Rev.}
\bvolume{162},
\bfpage{24}.
\doiurl{10.1016/j.earscirev.2016.09.004}.
\end{barticle}
\endbibitem


\bibitem[\protect\citeauthoryear{Schmalz and Stix}{1991}]{Schmalz1991}
\begin{barticle}
\bauthor{\bsnm{Schmalz}, \binits{S.}},
\bauthor{\bsnm{Stix}, \binits{M.}}:
\byear{1991},
\batitle{An alpha-Omega dynamo with order and chaos}.
\bjtitle{Astron. Astrophys.}
\bvolume{245},
\bfpage{654}.
\end{barticle}
\endbibitem





\bibitem[\protect\citeauthoryear{Schove}{1983}]{Schove1983}
\begin{bbook}
\bauthor{\bsnm{Schove}, \binits{D.J.}}:
\byear{1983},
\bbtitle{Sunspot Cycles},
\bpublisher{Hutchinson Ross Publishing Company}, \blocation{Stroudsburg}.
\end{bbook}
\endbibitem


\bibitem[\protect\citeauthoryear{Seilmayer
  \textit{et~al.}}{2012}]{Seilmayer2012}
\begin{barticle}
\bauthor{\bsnm{Seilmayer}, \binits{M.}},
\bauthor{\bsnm{Stefani}, \binits{F.}},
\bauthor{\bsnm{Gundrum}, \binits{T.}},
\bauthor{\bsnm{Weier}, \binits{T.}},
\bauthor{\bsnm{Gerbeth}, \binits{G.}},
\bauthor{\bsnm{Gellert}, \binits{M.}},
\bauthor{\bsnm{R{\"u}diger}, \binits{G.}}:
\byear{2012},
\batitle{Experimental evidence for a transient Tayler instability in a 
cylindrical liquid metal column}.
\bjtitle{Phys. Rev. Lett.}
\bvolume{108},
\bfpage{244501}.
\doiurl{10.1103/PhysRevLett.108.244501}.
\end{barticle}
\endbibitem




\bibitem[\protect\citeauthoryear{Sharp}{2013}]{Sharp2013}
\begin{barticle}
\bauthor{\bsnm{Sharp}, \binits{G.}}:
\byear{2013},
\batitle{Are Uranus and Neptune responsible for solar grand minima and solar cycle modulation?}
\bjtitle{Int J. Astron. Astrophys.}
\bvolume{3},
\bfpage{260}.
\doiurl{10.4236/ijaa.2013.33031}.
\end{barticle}
\endbibitem


\bibitem[\protect\citeauthoryear{Sokoloff and Nesme-Ribes}{1994}]{Sokoloff1994}
\begin{barticle}
\bauthor{\bsnm{Sokoloff}, \binits{D.}},
\bauthor{\bsnm{Nesme-Ribes}, \binits{E.}}:
\byear{1994},
\batitle{The Maunder minimum: a mixed-parity dynamo mode?}
\bjtitle{Astron. Astrophys.}
\bvolume{288},
\bfpage{293}.
\end{barticle}
\endbibitem


\bibitem[\protect\citeauthoryear{Solheim}{2013}]{Solheim2013}
\begin{barticle}
\bauthor{\bsnm{Solheim}, \binits{J.-E.}}:
\byear{2013},
\batitle{The sunspot cycle length - modulated by planets?}
\bjtitle{Pattern  Recogn. Phys.}
\bvolume{1},
\bfpage{159}.
\end{barticle}
\endbibitem






\bibitem[\protect\citeauthoryear{Soon \textit{et~al.}}{2014}]{Soon2014}
\begin{barticle}
\bauthor{\bsnm{Soon}, \binits{W.}},
\bauthor{\bsnm{Velasco Herrera}, \binits{V.M.}},
\bauthor{\bsnm{Selvaraj}, \binits{K.}},
\bauthor{\bsnm{Traversi}, \binits{R.}},
\bauthor{\bsnm{Usoskin}, \binits{I.}},
\bauthor{\bsnm{Chen}, \binits{C.A.}},
\bauthor{\bsnm{Lou}, \binits{J.Y.}}
\bauthor{\bsnm{Kao}, \binits{S.J.}},
\bauthor{\bsnm{Carter}, \binits{R.M.}},
\bauthor{\bsnm{Pipin}, \binits{V.}},
\bauthor{\bsnm{Severi}, \binits{M.}},
\bauthor{\bsnm{Becagli}, \binits{S.}}:
\byear{2014},
\batitle{A review of Holocene solar-linked climatic variation on centennial to
   millennial timescales: Physical processes, interpretative frameworks and
   a new multiple cross-wavelet transform algorithm}.
\bjtitle{Earth Sci. Rev.}
\bvolume{134},
\bfpage{1}.
\doiurl{10.1016/j.earscirev.2014.03.003}.
\end{barticle}
\endbibitem




\bibitem[\protect\citeauthoryear{Spruit}{2002}]{Spruit2002}
\begin{barticle}
\bauthor{\bsnm{Spruit}, \binits{H.}}:
\byear{2002},
\batitle{Dynamo action by differential rotation in a stably stratified stellar
  interior}.
\bjtitle{Astron. Astrophys.}
\bvolume{381},
\bfpage{923}.
\doiurl{10.1051/0004-6361:20011465}.
\end{barticle}
\endbibitem













\bibitem[\protect\citeauthoryear{Stefani \textit{et~al.}}{2016}]{Stefani2016}
\begin{barticle}
\bauthor{\bsnm{Stefani}, \binits{F.}},
\bauthor{\bsnm{Giesecke}, \binits{A.}},
\bauthor{\bsnm{Weber}, \binits{N.}},
\bauthor{\bsnm{Weier}, \binits{T.}}:
\byear{2016},
\batitle{Synchronized helicity oscillations: a link between planetary tides and the solar cycle?}
\bjtitle{Solar Phys.}
\bvolume{291},
\bfpage{2197}.
\doiurl{10.1007/s11207-016-0968-0}.
\end{barticle}
\endbibitem

\bibitem[\protect\citeauthoryear{Stefani \textit{et~al.}}{2017}]{Stefani2017}
\begin{barticle}
\bauthor{\bsnm{Stefani}, \binits{F.}},
\bauthor{\bsnm{Galindo}, \binits{V.}}
\bauthor{\bsnm{Giesecke}, \binits{A.}},
\bauthor{\bsnm{Weber}, \binits{N.}},
\bauthor{\bsnm{Weier}, \binits{T.}}:
\byear{2017},
\batitle{The Tayler instability at low magnetic Prandtl numbers: chiral 
symmetry breaking and synchronizable helicity oscillations}.
\bjtitle{Magnetohydrodynamics}
\bvolume{53},
\bfpage{169}.
\end{barticle}
\endbibitem


\bibitem[\protect\citeauthoryear{Stefani \textit{et~al.}}{2018}]{Stefani2018}
\begin{barticle}
\bauthor{\bsnm{Stefani}, \binits{F.}},
\bauthor{\bsnm{Giesecke}, \binits{A.}},
\bauthor{\bsnm{Weber}, \binits{N.}},
\bauthor{\bsnm{Weier}, \binits{T.}}:
\byear{2018},
\batitle{On the synchronizability of Tayler-Spruit and Babcock-Leighton type dynamos}.
\bjtitle{Solar Phys.}
\bvolume{293},
\bfpage{12}.
\doiurl{10.1007/s11207-017-1232-y}.
\end{barticle}
\endbibitem

\bibitem[\protect\citeauthoryear{Stefani, Giesecke and Weier}{2019}]{Stefani2019}
\begin{barticle}
\bauthor{\bsnm{Stefani}, \binits{F.}},
\bauthor{\bsnm{Giesecke}, \binits{A.}},
\bauthor{\bsnm{Weier}, \binits{T.}}:
\byear{2018},
\batitle{A model of a tidally synchronized solar dynamo}.
\bjtitle{Solar Phys.}
\bvolume{294},
\bfpage{60}.
\doiurl{10.1007/s11207-019-1447-1}.
\end{barticle}
\endbibitem


\bibitem[\protect\citeauthoryear{Stefani \textit{et~al.}}{2020a}]{Stefani2020a}
\begin{barticle}
\bauthor{\bsnm{Stefani}, \binits{F.}},
\bauthor{\bsnm{Giesecke}, \binits{A.}},
\bauthor{\bsnm{Seilmayer}, \binits{M.}},
\bauthor{\bsnm{Stepanov}, \binits{R.}},
\bauthor{\bsnm{Weier}, \binits{T.}}:
\byear{2020},
\batitle{Schwabe, Gleissberg, Suess-de Vries: Towards a consistent model of planetary synchronization of solar cycles}.
\bjtitle{Magnetohydrodynamics (in press)},
\bfpage{arxiv.org/abs/1910.10383}.
\end{barticle}
\endbibitem


\bibitem[\protect\citeauthoryear{Stefani \textit{et~al.}}{2020b}]{Stefani2020b}
\begin{barticle}
\bauthor{\bsnm{Stefani}, \binits{F.}},
\bauthor{\bsnm{Beer}, \binits{J.}},
\bauthor{\bsnm{Giesecke}, \binits{A.}},
\bauthor{\bsnm{Gloaguen}, \binits{T.}},
\bauthor{\bsnm{Seilmayer}, \binits{R.}},
\bauthor{\bsnm{Stepanov}, \binits{R.}},
\bauthor{\bsnm{Weier}, \binits{T.}}:
\byear{2020},
\batitle{Phase coherence and phase jumps in the Schwabe cycle}.
\bjtitle{Astron. Nachr. (submitted)},
\bfpage{arXiv:2004.10028}.
\end{barticle}
\endbibitem






\bibitem[\protect\citeauthoryear{Steinhilber \textit{et~al.}}{2012}]{Steinhilber2012}
\begin{barticle}
\bauthor{\bsnm{Steinhilber}, \binits{F.} \textit{et~al.}}:
\byear{2012},
\batitle{9,400 years of cosmic radiation and solar activity from ice cores and tree rings}.
\bjtitle{Proc. Natl. Acad. Sci.}
\bvolume{109},
\bfpage{5967}.
\doiurl{10.1073/pnas.1118965109}.
\end{barticle}
\endbibitem

\bibitem[\protect\citeauthoryear{Stepanov and Stefani}{2019}]{Stepanov2019}
\begin{barticle}
\bauthor{\bsnm{Stepanov}, \binits{R.}},
\bauthor{\bsnm{Stefani}, \binits{F.}}:
\byear{2019},
\batitle{Electromagnetic forcing of a flow with the azimuthal 
wave number m = 2 in cylindrical geometry}.
\bjtitle{Magnetohydrodynamics}
\bvolume{55},
\bfpage{207}.
\end{barticle}
\endbibitem



\bibitem[\protect\citeauthoryear{Takahashi}{1968}]{Takahashi1968}
\begin{barticle}
\bauthor{\bsnm{Takahashi}, \binits{K.}}:
\byear{1968},
\batitle{On the relation between the solar activity cycle and the solar tidal
  force induced by the planets}.
\bjtitle{Solar Phys.}
\bvolume{3},
\bfpage{598}.
\doiurl{10.1007/BF00151940}.
\end{barticle}
\endbibitem

\bibitem[\protect\citeauthoryear{Tayler}{1973}]{Tayler1973}
\begin{barticle}
\bauthor{\bsnm{Tayler}, \binits{R.J.}}:
\byear{1973},
\batitle{The adiabatic stability of stars containing magnetic fields-I: Toroidal fields}.
\bjtitle{Mon. Not. Roy. Astron. Soc.}
\bvolume{161},
\bfpage{365}.
\doiurl{10.1093/mnras/161.4.365}.
\end{barticle}
\endbibitem

\bibitem[\protect\citeauthoryear{Tlatov}{2013}]{Tlatov2013}
\begin{barticle}
\bauthor{\bsnm{Tlatov}, \binits{A.G.}}:
\byear{2013},
\batitle{Reversals of the Gnevyshev-Ohl rule}.
\bjtitle{Astrophys. J. Lett.}
\bvolume{772},
\bfpage{L30}.
\doiurl{10.1088/2041-8205/772/2/L30}.
\end{barticle}
\endbibitem



\bibitem[\protect\citeauthoryear{Usoskin \textit{et~al.}}{2007}]{Usoskin2007}
\begin{barticle}
\bauthor{\bsnm{Usoskin}, \binits{I.G.}},
\bauthor{\bsnm{Solanki}, \binits{S.K.}},
\bauthor{\bsnm{Kovaltsov}, \binits{G.A.}},
\byear{2007}
\batitle{Grand minima and maxima of solar activity: new observational constraints}.
\bjtitle{Astron. Astrophys.}
\bvolume{471},
\bfpage{301}.
\doiurl{10.1051/0004-6361:20077704 }.
\end{barticle}
\endbibitem


\bibitem[\protect\citeauthoryear{Usoskin \textit{et~al.}}{2016}]{Usoskin2016}
\begin{barticle}
\bauthor{\bsnm{Usoskin}, \binits{I.G.}},
\bauthor{\bsnm{Gallet}, \binits{Y.}},
\bauthor{\bsnm{Lopes}, \binits{F.}},
\bauthor{\bsnm{Kovaltsov}, \binits{G.A.}},
\bauthor{\bsnm{Hulot}, \binits{G.}},
\byear{2016}
\batitle{Solar activity during the Holocene: the Hallstatt cycle and its consequence for grand minima and maxima}.
\bjtitle{Astron. Astrophys.}
\bvolume{587},
\bfpage{A150}.
\doiurl{10.1051/0004-6361/201527295}.
\end{barticle}
\endbibitem

\bibitem[\protect\citeauthoryear{Vald\'es-Galicia and Velasco}{2008}]{Valdes2008}
\begin{barticle}
\bauthor{\bsnm{Vald\'es-Galicia}, \binits{J.F.}},
\bauthor{\bsnm{Velasco}, \binits{V.M.}},
\byear{2008}
\batitle{Variations of mid-term periodicities in solar activity physical phenomena}.
\bjtitle{Adv. Space. Res.}
\bvolume{41},
\bfpage{297}.
\doiurl{10.1016/j.asr.2007.02.012}.
\end{barticle}
\endbibitem









\bibitem[\protect\citeauthoryear{Vos \textit{et~al.}}{2004}]{Vos2004}
\begin{barticle}
\bauthor{\bsnm{Vos}, \binits{H.}},
\bauthor{\bsnm{Br\"uchmann}, \binits{C.}},
\bauthor{\bsnm{L\"ucke}, \binits{A.}},
\bauthor{\bsnm{Negendank}, \binits{J.F.W.}},
\bauthor{\bsnm{Schleser}, \binits{G.H.}},
\bauthor{\bsnm{Zolitschka}, \binits{B.}}:
\byear{2004},
\batitle{Phase stability of the solar Schwabe cycle in Lake 
Holzmaar, Germany, and GISP2, Greenland, between 10,000 and 9,000 cal. BP}
\bfpage{In: Fischer, H., Kumke, T., Lohmann, G., Fl\"oser, G., 
Miller, H., von Storch, H., Negendank, J. F. (Eds.), 
The Climate in Historical Times: Towards a Synthesis of 
Holocene Proxy Data and Climate Models, (GKSS School of Environmental 
Research), Springer, 293-318}.
\end{barticle}
\endbibitem

\bibitem[\protect\citeauthoryear{Weber \textit{et~al.}}{2013}]{Weber2013}
\begin{barticle}
\bauthor{\bsnm{Weber}, \binits{N.}},
\bauthor{\bsnm{Galindo}, \binits{V.}},
\bauthor{\bsnm{Stefani}, \binits{F.}},
\bauthor{\bsnm{Weier}, \binits{T.}},
\bauthor{\bsnm{Wondrak}, \binits{T.}}:
\byear{2013},
\batitle{Numerical simulation of the Tayler instability in liquid metals}.
\bjtitle{New J. Phys.}
\bvolume{15},
\bfpage{043034}.
\doiurl{10.1088/1367-2630/15/4/043034}.
\end{barticle}
\endbibitem

\bibitem[\protect\citeauthoryear{Weber \textit{et~al.}}{2015}]{Weber2015}
\begin{barticle}
\bauthor{\bsnm{Weber}, \binits{N.}},
\bauthor{\bsnm{Galindo}, \binits{V.}},
\bauthor{\bsnm{Stefani}, \binits{F.}},
\bauthor{\bsnm{Weier}, \binits{T.}}:
\byear{2015},
\batitle{The Tayler instability at low magnetic Prandtl numbers: between chiral
  symmetry breaking and helicity oscillations}.
\bjtitle{New J. Phys.}
\bvolume{17},
\bfpage{113013}.
\doiurl{10.1088/1367-2630/17/11/113013}.
\end{barticle}
\endbibitem

\bibitem[\protect\citeauthoryear{Weiss and Tobias}{2016}]{Weiss2016}
\begin{barticle}
\bauthor{\bsnm{Weiss}, \binits{N.O.}},
\bauthor{\bsnm{Tobias}, \binits{S.M}}:
\byear{2016},
\batitle{Supermodulation of the Sun's magnetic activity: the effects of
symmetry changes}.
\bjtitle{Mon. Not. Roy. Astron. Soc.}
\bvolume{456},
\bfpage{2654}.
\doiurl{10.1093/mnras/stv2769}.
\end{barticle}
\endbibitem




\bibitem[\protect\citeauthoryear{Wilmot-Smith
  \textit{et~al.}}{2006}]{Wilmotsmith2006}
\begin{barticle}
\bauthor{\bsnm{Wilmot-Smith}, \binits{A.L.}},
\bauthor{\bsnm{Nandy}, \binits{D.}},
\bauthor{\bsnm{Hornig}, \binits{G.}},
\bauthor{\bsnm{Martens}, \binits{P.C.H.}}:
\byear{2006},
\batitle{A time delay model for solar and stellar dynamos}.
\bjtitle{Astrophys. J.}
\bvolume{652},
\bfpage{696}.
\doiurl{10.1088/0004-637X/740/2/89}.
\end{barticle}
\endbibitem

\bibitem[\protect\citeauthoryear{Wilson}{1987}]{Wilson1987}
\begin{barticle}
\bauthor{\bsnm{Wilson}, \binits{R.M.}}:
\byear{1987},
\batitle{On the distribution of sunspot cycle periods}.
\bjtitle{J. Geophys. Res.}
\bvolume{92},
\bfpage{10101}.
\end{barticle}
\endbibitem


\bibitem[\protect\citeauthoryear{Wilson}{2008}]{Wilson2008}
\begin{barticle}
\bauthor{\bsnm{Wilson}, \binits{I.R.G.}}:
\byear{2008},
\batitle{Does a spin-orbit coupling between the Sun and the Jovian planets govern the solar cycle?}.
\bjtitle{Publ. Astron. Soc. Austr.}
\bvolume{25},
\bfpage{85}.
\doiurl{10.1071/AS06018}.
\end{barticle}
\endbibitem


\bibitem[\protect\citeauthoryear{Wilson}{2013}]{Wilson2013}
\begin{barticle}
\bauthor{\bsnm{Wilson}, \binits{I.R.G.}}:
\byear{2013},
\batitle{The Venus-Earth-Jupiter spin-orbit coupling model}.
\bjtitle{Pattern Recogn. Phys.}
\bvolume{1},
\bfpage{147}.
\doiurl{10.1086/508013}.
\end{barticle}
\endbibitem


\bibitem[\protect\citeauthoryear{Wolff and Patrone}{2010}]{Wolff2010}
\begin{barticle}
\bauthor{\bsnm{Wolff}, \binits{C.L.}},
\bauthor{\bsnm{Patrone}, \binits{P.N.}}:
\byear{2010},
\batitle{A new way that planets can affect the sun}.
\bjtitle{Solar Phys.}
\bvolume{266},
\bfpage{227}.
\doiurl{10.1007/s11207-010-9628-y}.
\end{barticle}
\endbibitem



\bibitem[\protect\citeauthoryear{Wood}{1972}]{Wood1972}
\begin{barticle}
\bauthor{\bsnm{Wood}, \binits{K.}}:
\byear{1972},
\batitle{Sunspots and planets}.
\bjtitle{Nature}
\bvolume{240}(\bissue{5376}),
\bfpage{91}.
\doiurl{10.1038/240091a0}.
\end{barticle}
\endbibitem




\bibitem[\protect\citeauthoryear{Zaqarashvili}{1997}]{Zaqa1997}
\begin{barticle}
\bauthor{\bsnm{Zaqarashvili}, \binits{T.}}:
\byear{1997},
\batitle{On a possible generation mechanism for the solar cycle.}.
\bjtitle{Astrophys. J.}
\bvolume{487},
\bfpage{930}.
\doiurl{10.1086/304629}.
\end{barticle}
\endbibitem




\bibitem[\protect\citeauthoryear{Zaqarashvili}{2018}]{Zaqarashvili2018}
\begin{barticle}
\bauthor{\bsnm{Zaqarashvili}, \binits{T.}}:
\byear{2018},
\batitle{Equatorial magnetohydrodynamic shallow water waves in the solar tachocline}.
\bjtitle{Astrophys. J.}
\bvolume{856},
\bfpage{32}.
\doiurl{10.3847/1538-4357/aab26f}.
\end{barticle}
\endbibitem



\end{thebibliography}
\end{document}